\documentclass[authoryear,preprint,review,12pt]{elsarticle}

\usepackage[utf8]{inputenc}
\DeclareUnicodeCharacter{2212}{-}
\usepackage{textcomp}
\usepackage{amssymb}
\usepackage{amsmath}
\usepackage{natbib}
\usepackage{float}
\usepackage{enumitem}
\newsavebox{\promptboxcontent}
\newenvironment{promptbox}[1][]{%
  \par\medskip\noindent\hrule\medskip\noindent\textbf{#1}\par\smallskip\noindent
}{%
  \par\smallskip\hrule\medskip
}
\newenvironment{examplebox}{%
  \par\smallskip\noindent\hspace{1em}\begin{minipage}{0.9\linewidth}
}{%
  \end{minipage}\par\smallskip
}
\newenvironment{exampleboxpurple}{%
  \par\smallskip\noindent\hspace{1em}\begin{minipage}{0.9\linewidth}
}{%
  \end{minipage}\par\smallskip
}
\usepackage{multicol}
\usepackage{tabularx}
\usepackage{longtable}
\usepackage{booktabs}

\usepackage{color}

\journal{International Journal of Human-Computer Studies}

\begin{document}

\begin{frontmatter}


\title{Conversational Decision Support for Information Search Under Uncertainty: Effects of Gist and Verbatim Feedback}

\author[uiuc]{Kexin Quan\corref{cor1}}
\ead{kq4@illinois.edu}

\author[uiuc]{Jessie Chin}
\ead{chin5@illinois.edu}

\cortext[cor1]{Corresponding author}

\affiliation[uiuc]{organization={Information Sciences, 
    University of Illinois Urbana-Champaign},
            country={USA}}

\begin{abstract}
Many real-world decisions rely on information search, where people sample evidence and decide when to stop under uncertainty. The uncertainty in the environment, particularly how diagnostic evidence is distributed, causes complexities in information search, further leading to suboptimal decision-making outcomes. Yet AI decision support often targets outcome optimization, and less is known about how to scaffold search without increasing cognitive load. We introduce SERA, an LLM-based assistant that provides either gist or verbatim feedback during search. Across two experiments ($N_{1}=54$, $N_{2}=54$), we examined decision-making outcomes and information search in SERA-Gist, SERA-Verbatim, and a no-feedback baseline across three environments varying in uncertainty. The uncertainty in environment is operationalized by the perceived gain of information across the course of sampling, which individuals may experience diminishing return of information gain (decremental; low-uncertainty), or a local drop of information gain (local optimum; medium-uncertainty), or no patterns in information gain (high-uncertainty), as they search more. Individuals show more accurate decision outcomes and are more confident with SERA support, especially under higher uncertainty. Gist feedback was associated with more efficient integration and showed a descriptive pattern of reduced oversampling, while verbatim feedback promoted more extensive exploration. These findings establish feedback representation as a design lever when search matters, motivating adaptive systems that match feedback granularity to uncertainty.
\end{abstract}


\end{frontmatter}


\section{Introduction}
\label{sec:introduction}

People today face an unprecedented volume of information and choices, and excessive decision-making can impose stress and cognitive strain \citep{baumeister1998, englert2021, levav2010, tran2019}. Classic work shows that information overload emerges when people must evaluate too many alternatives or attributes, degrading judgment quality \citep{malhotra1982, buchanan2001}. In contemporary digital settings, continuous notifications and competing content further tax attention and working memory, making it harder to evaluate evidence and sustain high-quality choices over time. As these demands accumulate, decision fatigue depletes cognitive resources and increases reliance on shortcuts \citep{levav2010, tran2019}.

AI-based decision support has become widely deployed to help users navigate these demands, from healthcare diagnostics \citep{rajpurkar2022ai} to financial planning \citep{berkel2023map} and everyday consumer choices \citep{shuai2023whoshouldi}. Yet effectiveness depends not only on predictive accuracy, but also on how systems communicate evidence and uncertainty in ways users can incorporate into their decisions \citep{xu2025confronting, lai2023empirical}. Prior work has largely emphasized outcome optimization, where systems provide recommendations and explanations intended to improve final decision quality \citep{lai2023empirical, arrieta2020explainable}. Explainable AI (XAI) methods, such as feature attributions, uncertainty estimates, and counterfactuals, aim to calibrate reliance by helping users assess model outputs \citep{arrieta2020explainable, bhatt2021uncertainty}. However, evidence of consistent benefit in complex decisions is mixed, and persuasive explanations can even increase overreliance \citep{cabitza2024never, bussone2015role, jacobs2021machine}. More fundamentally, these approaches are typically evaluated at the point of choice, with limited evidence on how feedback shapes information search during decision-making under uncertainty.

In many real-world decisions, descriptions about the decision environment are not available all at once. Instead, people need to continuously seek information to gauge the payoff structure of the decision environment and decide what to search next, how long to continue, and when they have seen enough to commit to a choice \citep{hertwig2004, hertwig2009, browne2004stopping}. Research on decisions from experience shows that choices depend on the information actually sampled, and that attentional constraints often lead people to stop early and underweight rare events \citep{hills2010}. These challenges are amplified under uncertainty in the environment, where decision makers must integrate partial observations of the environment continuously and adapt to the decision environment through the tradeoff between exploration (continuing searching new information) and exploitation (making decisions based on the sampled information). \citep{weber1987, cabrerizo2010}. We represent this uncertainty through the temporal distribution of diagnostic information, that is whether the perceived gain of information sampled follows a predictable pattern or appears irregularly across the course of information search.

Environmental structure directly constrains how well people can regulate exploration (searching more information) and exploitation (stopping to make decisions) \citep{guan2020threshold}. When perceived gain of information declines predictably, users can anticipate diminishing returns and apply stable stopping criteria \citep{todd2012ecological, gigerenzer2011}. When perceived gain of information fluctuates randomly or shows local optimum, the environment offers few reliable regularities and weak cues about whether continued information search will pay off \citep{browne2007cognitive, shadlen2016decision}. As a result, decision support should be expected to adapt to varying payoff structures of the environment, and support optimal regulation between exploration and exploitation. 
However, existing AI decision support rarely treats information search under uncertainty as a primary design variable, leaving unclear how feedback should adapt as predictability of the payoff structure of the environment varies \citep{lai2023empirical, arrieta2020explainable}.
We address this gap by examining how feedback representation interacts with environmental uncertainty in AI-assisted decision-making. We introduce SERA (Self-Regulatory Assistant), a large language model-based assistant that provides feedback as users search information. Drawing on Fuzzy-Trace Theory \citep{reyna2008theory, reyna1991fuzzy}, SERA delivers two forms of feedback: gist feedback compresses available evidence into a high-level summary that emphasizes what matters most at the current point in search, whereas verbatim feedback preserves detailed comparisons between items and attributes. We manipulate environmental uncertainty by varying the perceived gain of information encountered over time: (1) Low-uncertainty (Decremental): the perceived gain of information gradually declines when people search more, (2) medium-uncertainty (Local Optimum): the perceived gain of information would first show local drop followed by a rise to the maximum gain as people search more, and (3) high-uncertainty (random): the perceived gain of information appears in random patterns as people search more. The predictability of payoff structure of the decision environment favors different strategies to regulate exploration and exploitation. For the low-uncertainty condition, individuals would benefit by adopting a stopping rule to monitor the changes in information gain and stop searching when they meet the diminishing return \citep{charnov1976optimal}. For the medium-uncertainty environment, individuals would benefit by adopting a stopping rule to monitor more global changes in information gain to avoid insufficient exploration due to local optimum \citep{fu2006suboptimal}. The random environment would favor the strategy to stop based on the search cost \citep{seale2000optimal}. We evaluate both outcome effectiveness (decision accuracy, confidence) and process regulation (information search and stopping behavior) to characterize when feedback helps users regulate search and optimize decision outcome.

Our work offers three contributions:
\begin{itemize}
    \item A process-oriented account of feedback representation in decision-making under uncertainty, comparing gist and verbatim feedback across information environments that vary in predictability of the payoff structure.
    \item Evidence that environmental uncertainty moderates the value of decision support, clarifying when conversational feedback improves regulation of exploration and exploitation during decision-making.
    \item Design implications for adaptive systems that adjust feedback granularity to match environmental demands.
\end{itemize}

These contributions motivate three research questions:
\begin{itemize}
    \item RQ1 (Outcome effectiveness): How does SERA feedback affect decision accuracy and confidence across different levels of environmental uncertainty?
    \item RQ2 (Process regulation): How do gist versus verbatim representations shape information search and stopping behavior?
    \item RQ3 (Human-AI interaction): How do user perceptions and individual differences relate to reliance on SERA?
\end{itemize}

\section{Related Work}

\subsection{Information Search and Stopping Under Overload}
Human decision-making is adaptive in how it allocates effort across information search, integration, and stopping. Models of adaptive decision-making describe strategy selection as a tradeoff between accuracy and cognitive cost, where people shift along a continuum from information frugal approaches to more compensatory integration depending on task demands \citep{payne1993adaptive}. Empirical studies show that when memory load is high or information is structured predictably, people switch to simpler processing rules that minimize effort \citep{broder2003}. These shifts are visible in interaction traces, selective search concentrates on diagnostic cues while ignoring redundant information \citep{dhami2010}. These adaptive patterns become more consequential in information-rich environments, where cognitive limits constrain how much information users can process. Classic accounts of working memory and cognitive load argue that attention and memory capacity are bounded, and performance deteriorates when environments exceed these limits \citep{miller1956magical,sweller1988}. Empirical work on choice overload shows that larger assortments can reduce engagement and satisfaction \citep{iyengar2000choice,scheibehenne2010,chernev2015}. In digital settings, overload is amplified by fragmented presentation and constant updates, where communication overload and feature overload compete for attention and increase coordination costs \citep{karr2010more}. These works imply that decision difficulty is shaped not only by volume, but also by how information is structured and sequenced in the interface.

Decision-making under uncertainty makes this regulation problem explicit because users must decide how long to search as outcomes are learned through information search. Decision from experience work shows that choices depend on sampled information rather than full descriptions, and when attention is taxed, people often truncate search and underweight rare events, producing systematic distortions in risk beliefs and stopping behavior \citep{hertwig2004,hills2010}. Across sequential search and bandit paradigms, individuals adapt exploration and exploitation over time as uncertainty changes, and reduce redundant search as learning stabilizes \citep{hertwig2009,steyvers2009,schulz2018,gureckis2009}. These previous findings implicate the motivation that decision support in information-rich contexts should be framed as process regulation to help users allocate search effort and calibrate stopping as evidence accumulates.

\subsection{Decision Support Systems in Information Rich and Uncertain Tasks}
Decision support systems (DSS) have long been used to help people act in information-rich environments by aggregating data, highlighting patterns, and forecasting outcomes. A dominant line of AI-assisted decision-making operationalizes this as model prediction followed by an AI generated recommendation, the user then evaluates and accepts or rejects the advice \citep{lai2023empirical}. To strengthen this interaction, systems often attach additional support elements to predictions, including model explanations \citep{linardatos2020explainable} and uncertainty cues \citep{prabhudesai2023understanding, zhang2020effect, mendez2021showing}, with the intent of improving both decision quality and user acceptance in practice \citep{kawakami2022care}. While these designs can increase accuracy and efficiency in simplified or well structured tasks \citep{lai2020, vasconcelos2023explanations, yang2020visual}, they also expose persistent challenges in more complex decision processes, including low uptake of AI advice, difficulty integrating recommendations into domain reasoning, and overreliance on incorrect outputs \citep{mendez2021showing, bussone2015role, jacobs2021machine}. Importantly, evidence on whether explanations and related add-ons actually calibrate reliance is mixed, some studies report more appropriate reliance  \citep{wang2021explanations, yang2020visual, vasconcelos2023explanations}, but many show limited impact on accuracy \citep{cabitza2024never}, or even higher overreliance when explanations make advice feel more credible regardless of correctness \citep{bertrand2023questioning, bussone2015role, cabitza2024explanations, eiband2019impact, jacobs2021machine, schemmer2023appropriate}. This line of work frames DSS progress largely around improving end outcomes, by better predictions and better presentation, but it also indicates that interaction design, not only model quality, is central to whether AI support helps or harms.

A second strand examines reliance and appropriate use more directly, emphasizing that performance gains depend on users developing realistic expectations about when the system is likely to err \citep{mendez2021showing, bussone2015role}. This motivates decision support that shifts from single-shot prediction toward process-oriented assistance, especially in realistic settings where decisions rely on subjective and unstructured knowledge rather than a single objective label, such as healthcare, social welfare, and finance \citep{buccinca2022beyond, yang2023}. Instead of treating the recommendation as the ground truth solution, emerging work argues for AI that accompanies different phases of decision-making and augments the process \citep{zhang2024beyond}. Examples include evaluative support that helps users generate and assess hypotheses while leaving the decision to the user \citep{tim2023xai}, redesigns of clinical tools that aim to support the reasoning process leading up to a decision rather than pushing a conclusion \citep{yang2023}, and reflective approaches that present evidence for and against AI advice to prompt critical reflection \citep{cabitza2023let}. 
Empirical findings in applied domains further suggest that users often value AI for organizing evidence, surfacing constraints, and reducing tedious subtasks, rather than taking over interpretation \citep{yildirim2024multimodal, zhang2024beyond}. Even more restrained conversational approaches have explored decision support through context dependent probing questions that scaffold reasoning while minimizing the system’s role in analysis \citep{reicherts2022extending}. Across these designs, the key argument is that decision tasks occupy a spectrum of complexity, therefore the most suitable form of support can vary substantially, and direct recommendations are not always optimal \citep{ma2024beyond}.

Sequential decision settings introduce an additional layer, decisions unfold over multiple steps as users sample information and update beliefs over time. Early decision-support systems treat this as sequential information acquisition under cost and uncertainty, and formalize how a system can recommend which information source to query next and when to stop search \citep{moore1986model}. Later work in interactive information retrieval and decisions from experience similarly frames interface actions and search length as sequential decisions that can be optimized or learned over time \citep{zhang2016sequential, markant2015modeling}. These strands show that information acquisition and stopping can be modeled explicitly, yet most AI-assisted decision support still focuses on final prediction quality and keeps exploration–exploitation dynamics internal to the algorithm. It remains unclear how interface level feedback should help users regulate search effort and stopping in information-rich sequential tasks, a gap that motivates our focus on process regulation in such settings.

\subsection{Explainable Decision Support and Feedback Representation}

Explainable AI (XAI) has been introduced in recent years to address the opacity of modern machine learning models by providing human interpretable information about model behavior and prediction limitations \citep{arrieta2020explainable}. Post hoc methods were widely adopted to address the black-box concern by adding explanatory layers on top of trained models without changing their internal parameters. Existing XAI tools offer explanations in different forms, for example feature attributions that highlight influential input features \citep{arrieta2020explainable,hadash2022improving} and counterfactual explanations that describe minimal changes needed to alter a prediction \citep{karimi2020model}. These methods also operate at different levels of granularity, including local explanations for individual instances \citep{ribeiro2016should}, cohort or subgroup explanations for sets of related cases \citep{yuan2020subplex,cheng2020dece}, and global summaries that characterize model behavior across broader input spaces \citep{lundberg2014si}. The choice of form and granularity determines which aspects of the underlying decision logic become visible to users.

With growing emphasis on human-centered XAI, recent work studies explanations in terms of their function for users rather than only their algorithmic properties. Studies investigate how different explanation formats influence users’ mental models of a system, their trust and reliance, and their ability to detect and correct model errors \citep{arrieta2020explainable,raees2024explainable}. Findings indicate that effectiveness depends on task structure, expertise, and risk, and that no single explanation type works well across all contexts. This line of work motivates a shift from asking which explanation is correct in an abstract sense toward asking which feedback representation is appropriate for a given decision setting and decision role.

Conversational interfaces extend decision support beyond static panels into interactive dialogue where users can ask follow up questions, request clarification, and externalize partial reasoning. Conversational XAI applies this idea to AI-assisted decision-making by integrating explanations into large language model based dialogue. Prior work shows that conversational explanations can increase perceived understanding and trust, yet they also increase the risk of overreliance when fluent feedback is not well aligned with model reliability \citep{he2025conversational,gupta2022}. Existing evaluations in this space primarily focus on trust, reliance, and final task outcomes \citep{he2025conversational}, with limited attention to how conversational feedback should be represented to support regulation of information search in information-rich tasks \citep{raees2024explainable}.

Recent systems work further treats generative models as decision support tools that guide users through complex information spaces by questioning, summarizing, and reformulating content over time \citep{han2023redbot,reicherts2025ai,shuai2025towards}. In these systems, feedback is delivered as a sequence of conversational turns that can prompt reflection, help users monitor their uncertainty, and steer further search rather than as a single static explanation. However, empirical evidence remains limited on how specific representations of conversational feedback, for example compressed overviews versus detailed numeric traces, shape information search and stopping in decision under uncertainty. Representation level conversational feedback for decision support in information-rich, uncertain environments therefore remains an open area for our work's investigation.

\section{SERA System Design}
\label{sec:system-design}

\subsection{SERA Design Rationale}
We developed \textbf{SERA (Self-Regulatory Assistant)}, a conversational chatbot embedded in a decision-making task under uncertainty to support participants' ongoing sense-making and monitoring during decision-making. SERA allows participants to record key evidence and request on-demand summaries during information acquisition. The system was implemented as a web-based application (React.js frontend, Flask backend) and used an LLM to generate real-time summaries via predefined prompts. A Firebase database stored chat logs, behavioral traces, and interaction data for subsequent analysis.

\subsection{Interaction Flow and Self-Regulatory Scaffolding}
\begin{figure}[t]
    \centering
    \includegraphics[width=\linewidth]{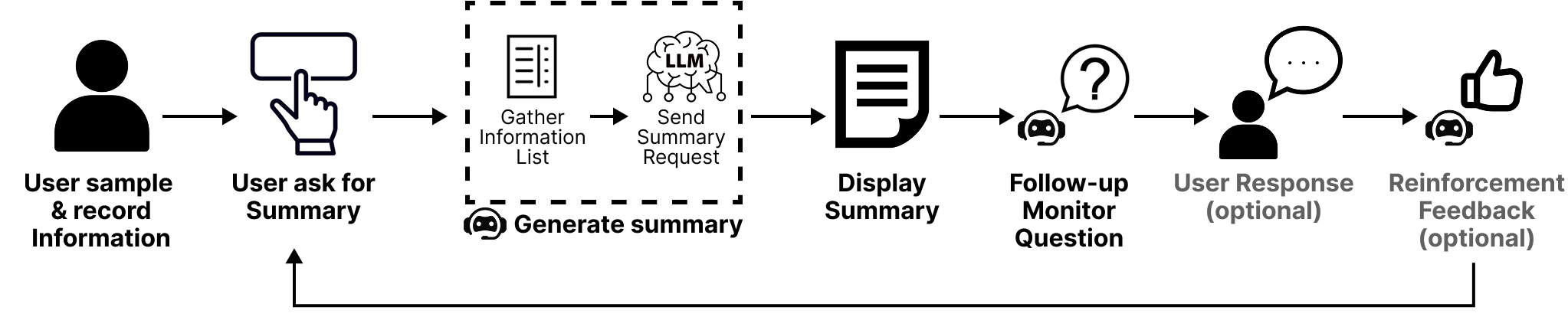}
    \caption{Iterative workflow of SERA during the decision-making process. 
   During information search, participants record key details and may request a summary. SERA provides summary feedback and prompts a monitoring question, after which participants can respond or continue search.}
    \label{fig:sera_workflow}
\end{figure}

SERA supports self-regulation by prompting participants to monitor their progress and reflect on remaining uncertainty during search (Figure~\ref{fig:sera_workflow}). Grounded in self-regulated learning theory \citep{zimmerman1986becoming}, SERA emphasizes \textit{sense-making} by helping participants organize the evidence they have recorded into concise summaries. After each summary, SERA presents a brief \textit{monitoring} prompt that asks participants to state their current priorities or what information they still need, before they continue search or stop to decide.

\subsubsection{Usage Scenario}
A participant begins a scenario with two options (A and B) displayed side by side. Each option contains hidden information pieces revealed one at a time through search under uncertainty. The participant clicks A or B to reveal a new piece of information, which is appended to the corresponding list. During searching, the participant may record brief notes in SERA's input box to capture evidence they consider salient. When needed, the participant requests a summary based on these notes and answers a short monitoring prompt about priorities or remaining uncertainty. The participant then continues search or stops and proceeds to the final choice between A and B.

\subsection{Feedback Representation in SERA}
To examine how feedback representation shapes integration of sequentially sampled evidence, SERA provides two summary formats \citep{brainerd2002}: \textbf{Gist} summaries, which emphasize essential meaning and core relationships between options, and \textbf{Verbatim} summaries, which preserve more literal details and specific attributes. In both conditions, SERA generates summaries from the same participant-entered notes, with representation format varying by condition.

Summaries were generated using a one-shot, role-based prompt that instructed the model to act as a neutral assistant for non-expert decision-makers. The prompt required outputs to be concise and impartial restatements of the participant's recorded notes. We used the OpenAI API (\texttt{gpt-4o-mini}) with fixed parameters (temperature $= 1$, maximum tokens $= 150$, top\_p $= 0$, presence\_penalty $= 0$) to maintain consistent behavior across feedback types. Full prompt instructions are provided in~\ref{appendix:summary_prompts}.

\subsubsection{Decision Task Paradigm: Search and Stopping}
\label{sec: core_paradigm}

A unified decision-making paradigm was applied that adapts the established searching-and-deciding framework from decision-from-experience research \citep{hertwig2004, hertwig2009}. 

\subsubsection{Two-Stage Decision Process}
Our experimental design follows a two-stage structure commonly used in decision-from-experience research \citep{camilleri2011, lejarraga2016}: \textbf{Search Stage.} Participants freely explore information from two decision options (A and B) without immediate consequences. Unlike traditional monetary gamble paradigms that present numerical outcomes, our adaptation used descriptive information pieces representing real-world decision attributes. Participants search information by clicking buttons corresponding to each option, each click revealing one piece of information. The search stage continued until participants felt ready to make their final decision. \textbf{Choice Stage.} After deciding they had sufficient information, participants proceeded to a final choice between the two options. This choice served as our primary dependent variable for decision accuracy.

\subsubsection{Information Gain and Ground Truth}
To bridge controlled laboratory paradigms with realistic decision contexts, each option comprised a list of descriptive information pieces representing key attributes of the scenario domain, rather than abstract numerical payoffs \citep{hertwig2009}. The pieces were crafted to be realistic and domain-appropriate, matched in length (15–19 words) to control for reading time. To manipulate the changes in perceived information gain (i.e., importance of information) during search, an estimated value was assigned to each piece of information indicating their expected contribution to decision outcomes. These values were validated through norming studies with independent participant samples (see~\ref{appendix:dm_scenarios}). 
Participants were unaware of the underlying importance values and the total number of available pieces. In all scenarios, one option was objectively superior based on aggregated importance-weighted scores to provide a ground truth for evaluating decision accuracy while maintaining the subjective complexity of real-world choices (see~\ref{appendix:findbestoption} for details).

\subsubsection{Interface for Behavioral Measurement}
\begin{figure}[t]
    \centering
    \includegraphics[width=\linewidth]{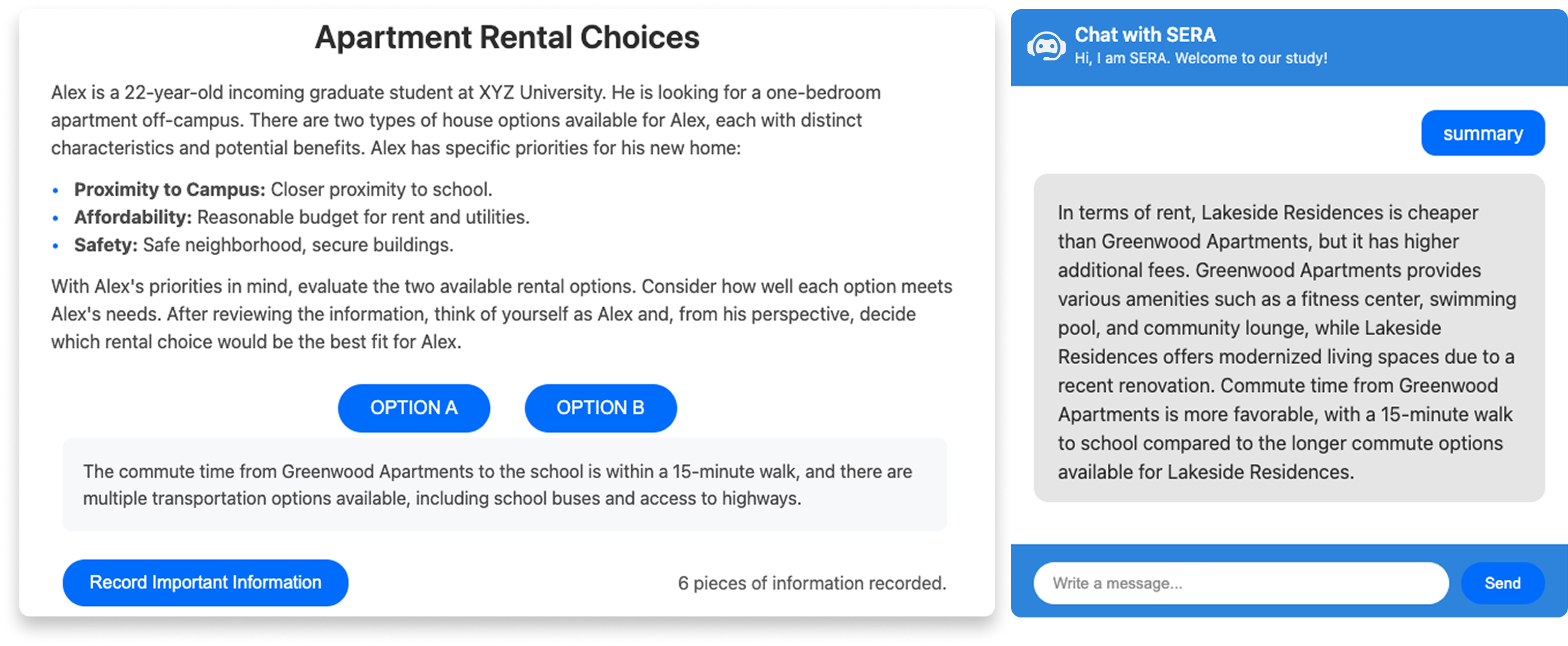}
    \caption{Example of study layout, with decision-making scenario (left) and SERA interface (right) showing summary results based on recorded information.}
    \label{fig: study1layout}
\end{figure}

The experimental interface was designed to support naturalistic decision-making while enabling precise behavioral measurement (Figure~\ref{fig: study1layout}). It employed a dual-panel layout where options A and B were displayed side by side, each with a sampling button and a cumulative list of sampled information. Information was disclosed progressively as participants sampled. 
The interface integrated SERA to provide summary assistance without disrupting search–decision workflow. Participants could type to request summaries at any point and respond to monitoring prompts before continuing exploration.  

\subsubsection{Environmental Uncertainty}
\label{sec:information_orders}
We operationalized environmental uncertainty through three conditions that vary in the predictability of change in perceived information gain across the course of search. Each distribution creates a distinct decision-making context and examines the effect of environments on how users regulate search:

\begin{itemize}
\item \textbf{Low-uncertainty, Decremental:} Information of highest perceived gain (i.e., importance) appeared first, followed by gradually decreasing perceived gain, creating a predictable declining pattern \citep{Rey2020}. Such environments resemble contexts where key information is immediately available, such as executive summaries or prioritized search results. As long as users experience the diminishing return in information gain, the stopping rule would be clear. 

\item \textbf{Medium-uncertainty, Local Optimum:} The perceived gain of information would first rise locally (with the moderately important information appearing first), then decline, and then peak at the most diagnostic content before declining again \citep{hills2010}. The resulting pattern creates a decision trap: users who stop after the initial peak miss the most valuable information later in the sequence. These local-optimum conditions could elicit complexity in search under uncertainty, where insufficient exploration would lead to suboptimal performance.

\item \textbf{High-uncertainty, Random:} Information importance was randomized, making perceived gain of information completely unpredictable throughout the course of search. In these high-uncertainty environments, the estimated gain of continued search cannot be inferred from prior samples. Without reliable cues for when to stop, participants in the random condition face the greatest challenge in regulating their search, making it the most likely setting to reveal benefits of external scaffolding.
\end{itemize}

\subsection{Experimental Design}
\label{sec:general_design}
We implemented a 3×3 mixed-factorial design with feedback representation as a between-subjects factor and environmental uncertainty as a within-subjects factor. This design maximizes statistical power while controlling for individual differences in decision-making styles and abilities.

\subsubsection{Between-Subjects Factor: Feedback Representation}
We manipulated the representation of information to provide different types of feedback facilitating participants' integration and uptake of multiple information pieces. Participants were randomly assigned to one of three conditions: (1) Control (No-SERA), in which decisions were made without assistance; (2) SERA-Gist (SERA-G), which provided gist summaries emphasizing essential meaning and core relationships between options; and (3) SERA-Verbatim (SERA-V), which provided verbatim summaries preserving literal details and specific attributes. Random assignment ensured balanced group sizes and minimized selection bias.

\subsubsection{Within-Subjects Factor: Environmental Uncertainty}
We manipulated environmental uncertainty by varying the perceived gain of information encountered over time. Each participant experienced all three distribution conditions in a counterbalanced order. The experiment followed a 3 (feedback representation: between-subjects) × 3 (environmental uncertainty: within-subjects) mixed-factorial design, yielding nine total conditions.

\subsection{Data Collection and Measures}

\subsubsection{Behavioral Data Capture}
The paradigm enabled comprehensive measurement of decision-making behavior. All \textit{search actions} were logged with precise timestamps, allowing analysis of search frequency, sequence, switching patterns, and temporal dynamics. \textit{SERA interaction logs}, including summary requests, generated outputs, monitoring questions, and participant responses, were also recorded to examine how AI assistance shaped subsequent search. In addition, \textit{key decision process metrics} were captured, including total search time, number of information pieces sampled per option, decision confidence, and final choice accuracy.   

\subsubsection{Common Questionnaires}
Standardized questionnaires were administered across both studies to assess individual differences, track participants’ decision-making experiences, and validate the experimental manipulations. Instruments were distributed before, during, and after the experiment, with attention checks embedded to ensure engagement.  

\textbf{Pre-Study Questionnaire.} At the outset, participants completed a demographic survey (age, gender, ethnicity, education). To capture individual differences, we included the Big Five Inventory-10 (BFI-10) to measure personality traits \citep{rammstedt2007}, and the General Decision-making Style (GDMS) questionnaire to assess preferences for rational, intuitive, dependent, avoidant, and spontaneous decision-making approaches \citep{scott1995}. 
Full details are provided in~\ref{appendix: pre-survey}.  

\textbf{Checkpoint Survey.} After each decision-making task, participants completed a short three-item survey designed to evaluate information comprehension and integration based on their search experience. Items used 7-point Likert scales (1 = “Strongly Disagree” to 7 = “Strongly Agree”) to assess (1) confidence in their choice \citep{bandura1997, zimmerman2000}, (2) ease of understanding the information \citep{sweller1988}, and (3) ability to synthesize and integrate across information pieces. Item details are provided in~\ref{appendix: ckpt_survey}.  

\textbf{Post-Study Questionnaire.} After completing the experiment, all participants completed a brief post-study survey. Participants in the SERA assisted conditions answered the Unified Theory of Acceptance and Use of Technology (UTAUT) scale \citep{venkatesh2003} to rate perceived usefulness, ease of use, and intention to adopt SERA. All participants also received optional open ended questions inviting them to reflect on their overall experience with the task and the system. Full item lists are provided in~\ref{appendix:post-survey}.

\subsubsection{Norming Study} 
Before the experimental tasks, participants completed a norming study to establish a behavioral baseline and filter out atypical decision-making. We adopted a card-draw paradigm \citep{hertwig2009} with two options (A and B), presented through the same interface as the main experiment but without AI assistance (Figure~\ref{fig:screening}). 
Option A yielded a 22\% chance of 10.5 points and a 78\% chance of 3.8 points (expected value = 5.46). Option B yielded a 30\% chance of 3.5 points and a 70\% chance of 2.5 points (expected value = 2.8). Thus, Option A was objectively superior. Participants could freely sample from either option before making a final choice by clicking “make your final choice.” Only those who selected the superior option advanced to the experimental tasks (Figure~\ref{fig:screening_choice}).

\section{Preliminary Study}
\label{sec:study1}

We conducted a preliminary study with three objectives: (1) to verify the feasibility of SERA as an embedded decision support tool, (2) to detect initial effect patterns that would inform hypotheses for the main study, and (3) to identify methodological limitations requiring refinement. This study used realistic decision scenarios (apartment renting, stock investment, medication selection) to examine whether SERA's gist and verbatim feedback mechanisms could influence decision-making behavior in ecologically valid contexts. We report key findings that motivated the main study design, rather than treating this as a standalone confirmatory study.

\subsection{Participants and Recruitment}
\label{sec:study1_participants}
We recruited 70 participants through Prolific. After excluding 16 participants who failed attention checks or had unrealistically short decision times below 30 seconds, the final sample consisted of 54 participants (36 female, 18 male; ages 18–65, $M = 31.40$, $SD = 9.10$). This sample size provided .80 power to detect a medium effect ($f = 0.25$, $\alpha = .05$) in the 3 × 3 mixed-factorial design. Random assignment resulted in 22 participants in SERA-G, 15 in SERA-V, and 17 in No-SERA conditions. Participants represented diverse ethnic backgrounds: 31 White or Caucasian, 10 Hispanic or Latino, 8 Black or African American, 4 Asian, and 1 Native American/Alaska Native. Most participants (81.5\%) had completed or were pursuing higher education. All participants were U.S.-based native English speakers. The study received IRB approval, and participants were compensated \$8 for their participation.   

\subsection{Study Procedure}
\label{sec:study1_procedure}
\begin{figure}[t]
    \centering
    \includegraphics[width=1.03\linewidth]{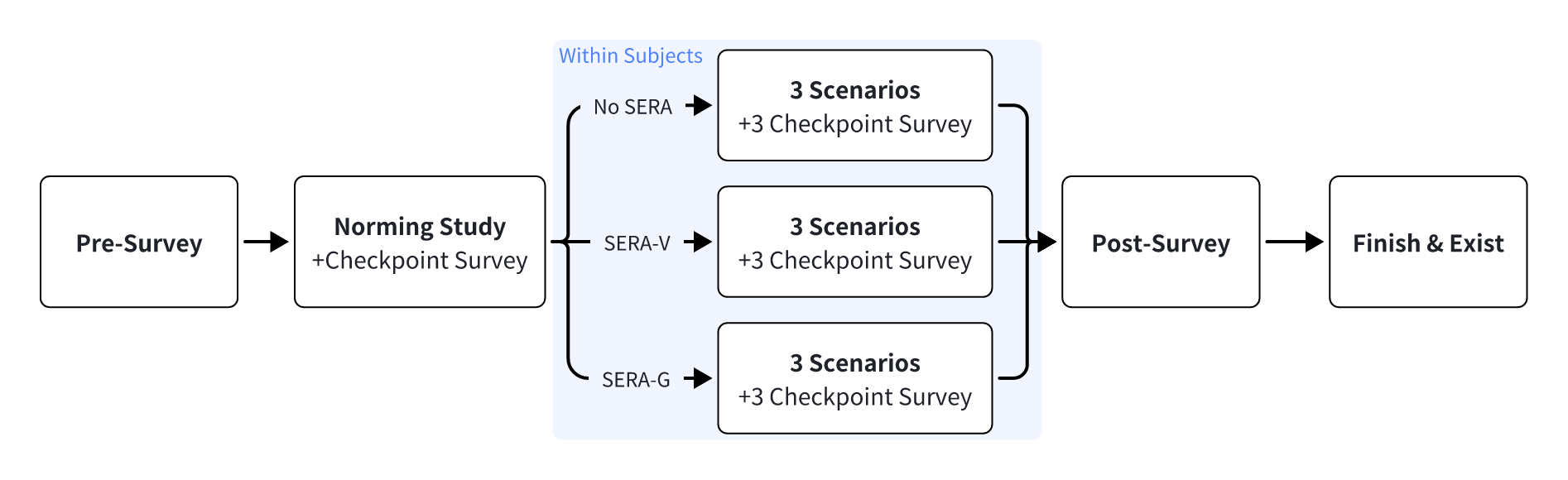}
    \caption{Preliminary study experimental flow.}
    \label{fig:study1design}
\end{figure}

The experimental session was conducted asynchronously online (see Figure~\ref{fig:study1design}).
Each session began with online informed consent and a pre-survey questionnaire (5 min), followed by a norming study (3 min). Participants who passed the norming study completed three decision-making scenarios presented in counterbalanced order, with a checkpoint survey administered after each scenario (8-10 min per scenario). The session concluded with a post-study questionnaire evaluating participants’ experiences with or without SERA (10 min).

\subsection{Preliminary Results}
\label{sec:study1_results}

\subsubsection{RQ1: SERA Improves Decision Outcomes Compared to No-SERA}

\paragraph{\textbf{SERA Improved Decision Accuracy and Confidence}} 
Mixed-effects models were conducted using the \texttt{lme4} and \texttt{lmerTest} packages in R. Condition (contrast 1: Control vs. SERA; contrast 2: SERA-G vs. SERA-V) and environmental uncertainty (contrast 1: others vs. Random; contrast 2: Decremental vs. Local-Opt) were entered as fixed effects, with random intercepts for participants and scenarios. Results showed significant two-way interactions between condition and environmental uncertainty (Condition C1 × Distribution C1: $B = -0.24$, $SE = 0.10$, $t = -2.37$, $p < .05$; Condition C2 × Distribution C2: $B = -0.33$, $SE = 0.13$, $t = -2.50$, $p < .05$). Participants achieved higher decision accuracy with SERA assistance than in the Control condition, particularly when information was presented randomly. In more structured sequences (Decremental), accuracy was high across all conditions. No significant accuracy differences emerged between SERA-G and SERA-V.

\subsubsection{RQ2: Effects of SERA Type and Environmental Uncertainty on Decision-making Behavior}
\label{sec:study1_rq2_rq3}

\paragraph{\textbf{Gist Was Associated with Less Oversampling Compared to Verbatim}}
To examine how SERA feedback type and environmental uncertainty influenced optimal stopping behavior, we analyzed whether participants sampled within, below, or above the predefined optimal information range using chi-square tests. The definition and derivation of this optimal range are detailed in~\ref{appendix:optimalrange}. Figure~\ref{fig:h1c} illustrates the distribution of stopping patterns across experimental conditions.

\begin{itemize}
    \item \textbf{Effect of SERA feedback}
Stopping patterns varied across SERA feedback types depending on environmental uncertainty. In the Decremental condition, SERA-G and SERA-V showed similar distributions across search categories, $\chi^{2}(2) = 1.03$, $p = .598$, $V = 0.17$. Under the Local-Opt distribution, SERA-G participants more often stopped within the optimal range (31.8\%), whereas SERA-V participants predominantly oversampled (80.0\%), $\chi^{2}(2) = 3.80$, $p = .150$, $V = 0.32$. Although this difference did not reach statistical significance, the pattern is consistent with gist-oriented feedback supporting more calibrated stopping when search incentives are prolonged.
    \item \textbf{Effect of environmental uncertainty}  
Stopping category distributions were comparable between Decremental and Local-Opt conditions, $\chi^{2}(2) = 0.72$, $p = .698$, $V = 0.08$. Descriptively, the Decremental distribution was associated with more calibrated stopping, with a higher proportion of participants stopping within or below the optimal range (44.4\%) than in Local-Opt (38.9\%), indicating that structured information presentation may help participants better judge when sufficient information has been acquired. 
\end{itemize}

\begin{figure}[t]
    \centering
    \includegraphics[width=\linewidth]{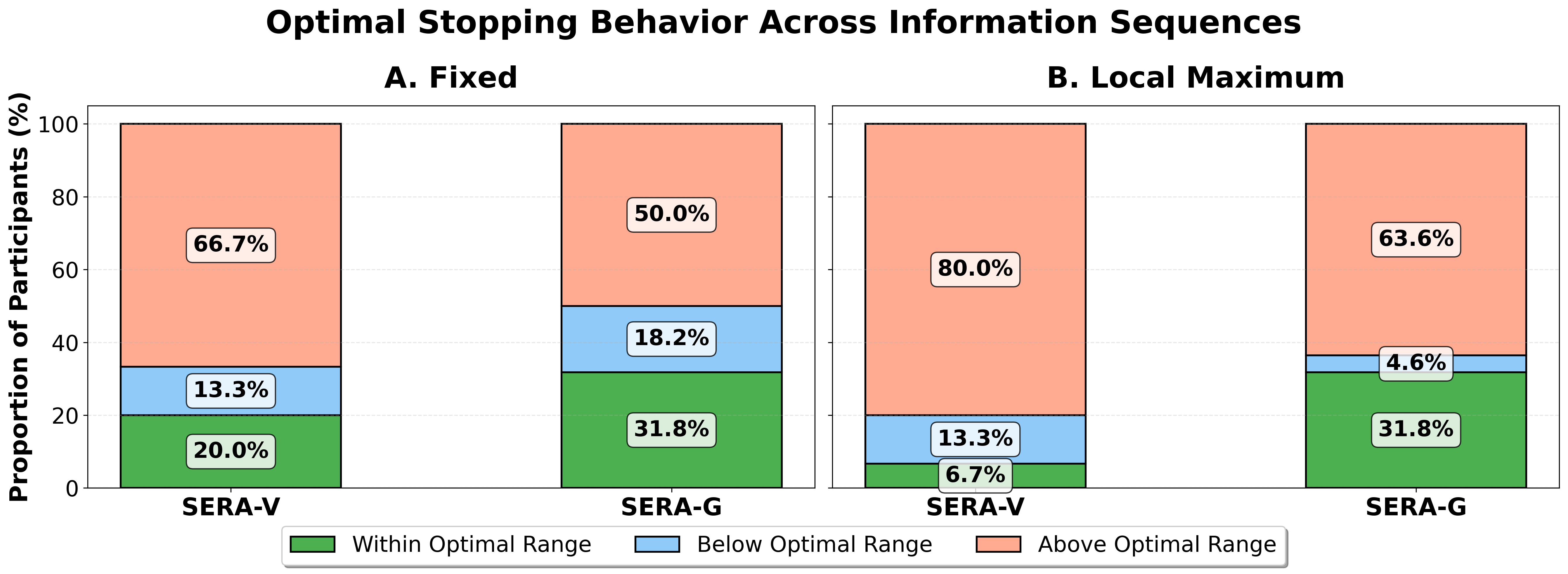}
    \caption{Preliminary Study: Distribution of stopping behavior relative to optimal information range across SERA feedback conditions.}
    \label{fig:h1c}
\end{figure}

\paragraph{\textbf{SERA-G Increased Early-Stage Switching}}
We examined exploratory behavior via mixed-effects models predicting switching rates, with agent condition (SERA-G, SERA-V, Control) and environmental uncertainty (Decremental, Local-Opt, Random) as fixed effects and participants and scenarios as random effects. The main effect of agent condition on total switching was not significant ($B = 0.86$, $SE = 1.39$, $t = 0.62$, $p = .54$). Planned contrasts showed higher switching for SERA-G than control ($B = 2.48$, $SE = 1.10$, $t = 2.25$, $p < .05$), with no difference between SERA-V and control ($B = 0.74$, $SE = 1.22$, $t = 0.60$, $p = .55$). No effects of information order or interactions were observed. To further examine temporal dynamics, we conducted a secondary analysis focusing on the first third of the decision period. Participants in SERA conditions exhibited higher switching rates during this early phase compared to control ($B = 0.79$, $SE = 0.38$, $t = 2.04$, $p < .05$), suggesting SERA feedback fosters early-stage exploratory engagement before decision convergence.

\paragraph{\textbf{Environmental Uncertainty Promoted Exploration}}
To characterize participants’ information search strategies, we analyzed search behavior during the initial exploration phase by examining the first third of sampled information pieces (or all samples if fewer than six). This early search window captured exploratory search patterns prior to decision convergence. Two independent coders, blind to study conditions, visually inspected temporal plots of participants’ option choices over time and classified each search pattern as \textit{Comprehensive} (systematic comparison through frequent switching between options) or \textit{Piecewise} (sequential search focused on one option before switching), following established methods \citep{hills2010} (Cohen’s $\kappa = 0.86$). Search patterns showed modest variation across decision scenarios, $\chi^{2}(2, N = 162) = 4.77$, $p = .092$, $V = .17$. 
Descriptively, the Decremental condition was associated with lower rates of piecewise sampling (14.8\%) compared to Local-Opt (29.6\%) and Random (31.5\%) distributions, aligning with more comprehensive comparative exploration. A similar pattern emerged across SERA feedback types, $\chi^{2}(2, N = 54) = 1.24$, $p = .537$, $V = .15$, with SERA-G participants showing higher piecewise sampling (31.8\%) than No-SERA (17.6\%) and SERA-V (20.0\%).

\subsubsection{RQ3: Individual Differences and User–System Interaction}
The preliminary study provided an initial exploration of RQ3, focusing on how individual traits and subjective perceptions shaped engagement with SERA. 

\paragraph{\textbf{Personality Traits and Decision Styles}}
Pearson correlations were conducted between personality traits and decision-making styles with behavioral measures: total number of SERA interactions, average switch rate between options, and average number of search actions. Two significant associations emerged. Rational decision-making style was positively correlated with SERA interactions ($r = .203$, $p = .028$), and Avoidant style showed a similar association ($r = .207$, $p = .025$). It suggests that participants who typically rely on structured analysis or who feel uncertain in complex contexts were both more likely to engage with SERA for support. No other correlations reached significance.

\paragraph{\textbf{Subjective Perceptions of SERA}}
Interview responses revealed divergent perceptions of SERA’s usefulness and trustworthiness. Many participants emphasized that \textbf{SERA alleviated decision-making} by organizing large amounts of information into concise comparative summaries. Participants described SERA as “sorting and managing information” and “presenting only the most relevant data” (P16, P18, P19), which helped them filter redundant details and focus on key differences. Several noted that SERA also \textbf{bridged knowledge gaps} in complex domains such as finance or medicine by transforming technical materials into digestible insights (P5, P11). These accounts illustrate SERA’s role as both a cognitive aid and a knowledge support tool that reduced cognitive strain and increased decision confidence.
Conversely, some participants preferred to rely on their own reasoning, citing either confidence in their decision strategies or skepticism about SERA’s impartiality. As one participant noted, “I hardly used SERA… the facts it provides can be skewed and biased” (P25). Others explained that they could process the information independently or preferred intuitive, rapid judgments without AI mediation (P4, P7, P10). These perspectives highlight that trust, perceived competence, and task difficulty jointly influence willingness to engage with AI support.

\subsection{Preliminary Study Discussion}
\label{sec:study1_discussion}

\subsubsection{Key Preliminary Findings}
The preliminary study provided initial evidence for SERA as a decision support tool in information overload contexts. 
First, SERA improved decision accuracy compared to no assistance, with the largest benefits under random information sequences where uncertainty was greatest. This pattern is consistent with prior work showing that unstructured environments impose higher demands on search and increase error rates \citep{hills2010}. 
Second, analyses of efficiency measures revealed that SERA increased decision time and information search, especially in the verbatim condition, where participants frequently engaged in oversampling. While gist summaries appeared to moderate this tendency by supporting more adaptive switching, verbatim feedback promoted more exhaustive exploration. 
Third, participants’ interaction patterns revealed frequent use of summary requests and monitoring responses, indicating that SERA not only provided decision support but also structured how participants alternated between information search and reflection.
These patterns motivated a stronger focus on information search and stopping behavior in the main study.

\subsubsection{Preliminary Study Limitations}
As a preliminary study, this work has several limitations that inform future refinement. First, the task may not have induced sufficient cognitive demand. Many participants described the search process as easy and low in stress, reducing the perceived need for SERA. To enhance ecological validity, main study designs should increase information volume and complexity, remove salient priority cues, and introduce explicit costs or outcomes to heighten cognitive demand. Second, participants with prior domain knowledge often relied on personal expertise rather than AI support, suggesting that SERA’s utility may vary with users’ familiarity and confidence. Third, the uncontrolled online setting may have reduced engagement and data quality. These limitations underscore the need for more challenging, ecologically valid tasks, controlled experimental conditions, and closer alignment with users’ knowledge levels to more accurately assess SERA’s contribution to decision-making.

\section{Main Study}
\label{sec: study2}

The main study served as the confirmatory test of hypotheses generated in the preliminary study. We addressed three methodological limitations identified earlier and tested whether the initial effect patterns would replicate under more controlled conditions.

\subsection{Study Design Improvements}
\label{sec:study2_design}
Building on the preliminary findings, the main study incorporated three refinements to enhance experimental validity and control. First, to eliminate prior knowledge confounds, realistic domains were replaced with hypothetical scenarios, and explicit priority cues were removed from descriptions (see~\ref{appendix:study2infodesign}). This design ensured participants relied solely on the provided information to determine value.

Second, to standardize the information environment, each option was shortened to 25 items based on preliminary study search patterns. Six information lists were collaboratively constructed and independently coded by two researchers, yielding good reliability ($r = 0.68$).

Third, to increase cognitive demand and strategic engagement, a points-based incentive system was implemented. Participants began each scenario with 120 points, incurring a cost of 3 points per sample, earning 6 bonus points for identifying the top three most valuable items per option, and 50 points for selecting the objectively superior option.

\subsection{Participants and Recruitment}
We recruited 54 participants through the university email list. This sample size was determined by an a priori power analysis for the 3 × 3  mixed-factorial design to detect a medium effect ($f = 0.25$, $\alpha = .05$, power = .80). Participants ranged in age from 18 to 64, with the majority between 18–34. The sample comprised 31 females, 21 males, and 2 agender participants, with diverse ethnic backgrounds, predominantly Asian (60\%) and White (37\%). Most participants (94\%) were pursuing or had completed higher education, and about 75\% were native English speakers. The study received IRB approval. Compensation was performance-based, ranging from \$9 to \$13 for approximately one hour of participation.

\subsection{Main Study Procedure}

\begin{figure}[t]
    \centering
    \includegraphics[width=1.03\linewidth]{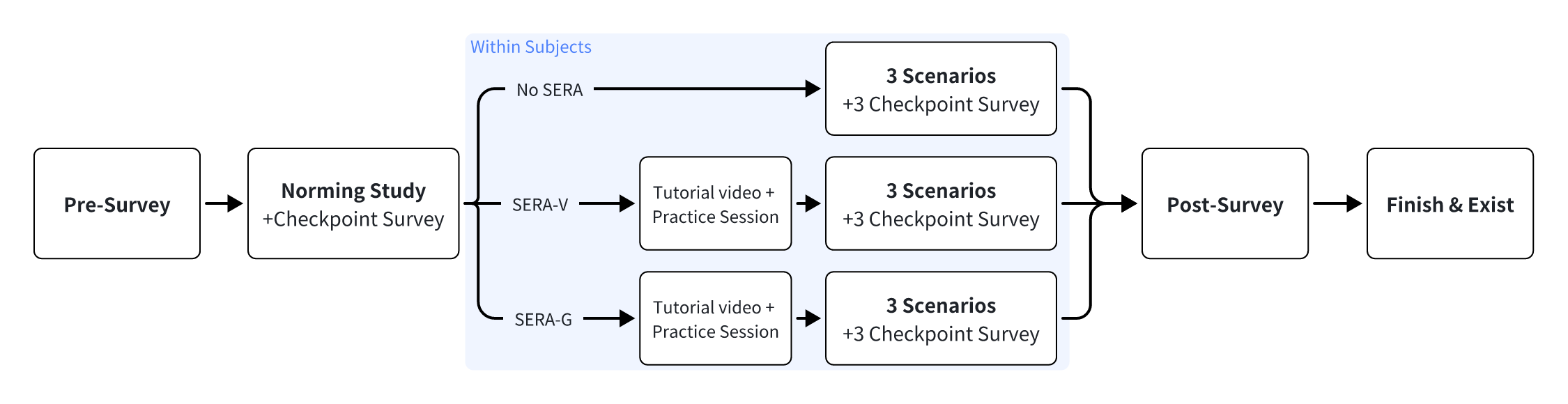}
    \caption{Main study experimental flow.}
    \label{fig:study2design}
\end{figure}

The main study employed supervised data collection with screen sharing to ensure high data quality and adherence to protocol. As illustrated in Figure~\ref{fig:study2design}, each session began with informed consent followed by a pre-survey questionnaire (5 min). Participants completed a brief norming study (3 min) with a checkpoint survey to establish baseline performance.  
Participants were assigned to all three within-subject conditions presented in counterbalanced order. In SERA conditions, participants first viewed a tutorial video and completed a practice session before proceeding. For each condition, participants worked through three decision-making scenarios, with a checkpoint survey administered after each scenario (8-10 min per scenario).  
The study concluded with a post-survey (10 min) assessing participants’ overall experience with and without SERA. Throughout the session, the researcher monitored participants via screen sharing to maintain engagement and ensure compliance with the study protocol.

\subsection{Main Study Results}
To examine the fixed effects of three agent conditions (control, SERA-G, and SERA-V) and environmental uncertainty (Decremental, Local-Opt, and Random) on decision-making outcomes and information search behavior, we conducted mixed-effects models using R packages lme4 and lmerTest. We chose mixed-effects modeling to consider the variance of participants and decision-making scenarios as random effects. We used contrast coding to examine the effects of categorical variables. For the effects of agent condition, we used two contrasts to compare the differences between the control and two SERA conditions (Contrast 1), and the differences between SERA-G and SERA-V (Contrast 2). For the effects of environmental uncertainty, we used two contrasts to compare the differences between the non-random and random conditions (Contrast 1), and the differences between the decremental and local-opt conditions (Contrast 2).

\subsubsection{RQ1: SERA Improves Decision Outcomes Compared to No-SERA}
\label{main h1 results}

\begin{figure}[ht]
    \centering
    \includegraphics[width=\linewidth]{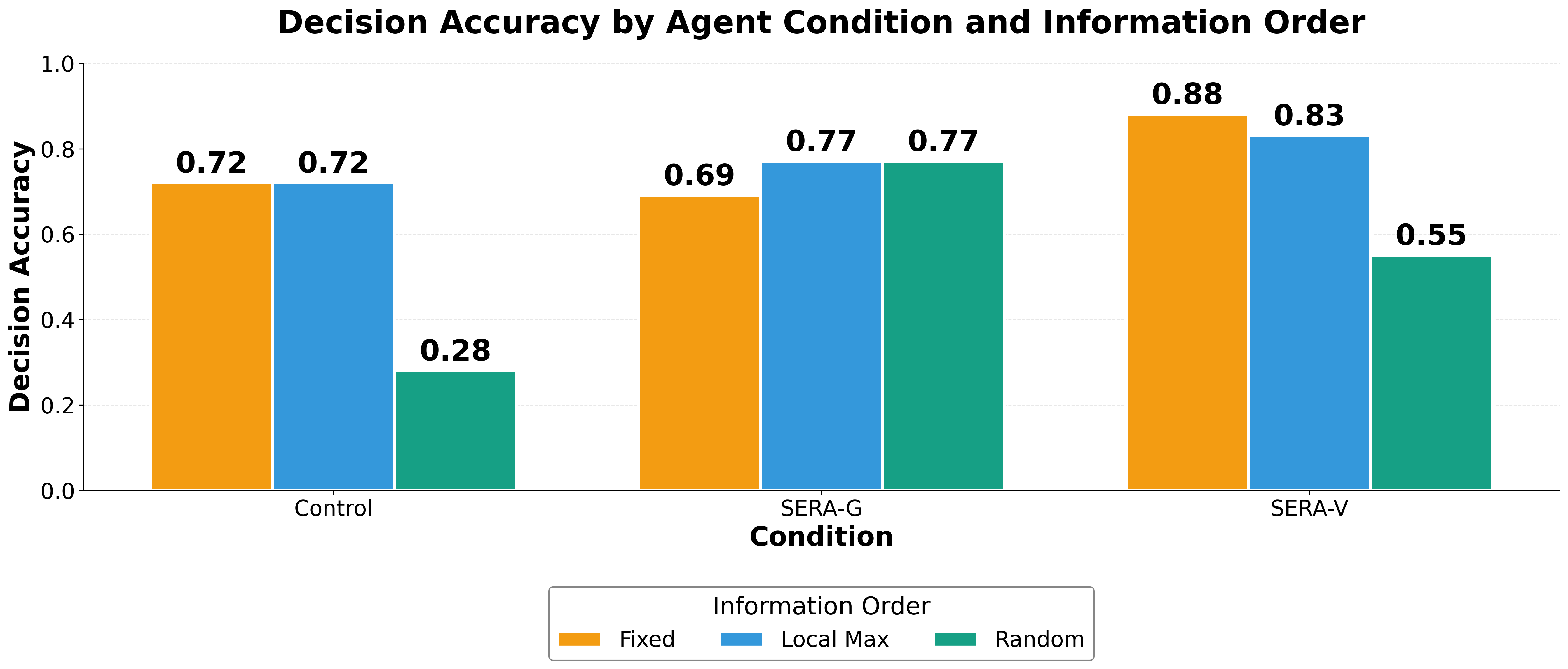}
    \caption{Decision accuracy by agent condition and environmental uncertainty.}
    \label{fig:decisionaccuracy}
\end{figure}

\paragraph{\textbf{Participants Made More Accurate Decisions with SERA}}
We examined the effects of agents and environmental uncertainty on decision-making performance (i.e., accuracy of the choice), considering random effects of participants and scenarios (see Model 1 in Table~\ref{tab:decisionaccuracy}). We found significant main effects of agent condition ($B=0.18$, $SE=0.07$, $t=2.42$, $p<.05$) and environmental uncertainty ($B=0.24$, $SE=0.07$, $t=3.72$, $p<.05$), showing that people made more accurate decisions with SERA compared to the control condition, and made more accurate decisions when the information was presented in structured distributions than in the random distribution. Findings also showed the significant interaction effects of agents and environmental uncertainty, suggesting participants made more accurate decisions with SERA than the control condition, especially when the information was presented randomly ($B=-0.31$, $SE=0.14$, $t=-2.24$, $p<.05$); which effects were even larger when participants used SERA-G than SERA-V ($B=-0.34$, $SE=0.16$, $t=-2.10$, $p<.05$) (see Figure~\ref{fig:decisionaccuracy}). Hence, participants took more advantage of SERA-G to make their decisions in demanding contexts, especially with uncertainty, compared to other conditions. 

\begin{table}[ht]
\centering
\caption{Mixed-effects model predicting decision accuracy by agent condition and environmental uncertainty (Model 1).}
\label{tab:decisionaccuracy}
\begin{tabular}{lcc}
\toprule
\textbf{Predictor} & \textbf{B (SE)} & \textbf{t} \\
\midrule
Intercept & 0.69 (0.09) & 7.57** \\
Agent 1. Control vs SERA & 0.18 (0.07) & 2.42* \\
Agent 2. SERA-G vs SERA-V & -0.01 (0.09) & -0.17 \\
Environmental Uncertainty 1. Non-Random vs Random & 0.24 (0.07) & 3.72** \\
Environmental Uncertainty 2. Decremental vs Local-Opt & 0.01 (0.07) & 0.10 \\
Agent 1 $\times$ Environmental Uncertainty 1 & -0.31 (0.14) & -2.24* \\
Agent 2 $\times$ Environmental Uncertainty 1 & -0.34 (0.16) & -2.10* \\
Agent 1 $\times$ Environmental Uncertainty 2 & 0.01 (0.16) & 0.07 \\
Agent 2 $\times$ Environmental Uncertainty 2 & 0.13 (0.18) & 0.72 \\
\midrule
Log Likelihood & \multicolumn{2}{c}{-83.7} \\
\bottomrule
\multicolumn{3}{l}{\textit{Note.} *$p<.05$; **$p<.01$}
\end{tabular}
\end{table}

\paragraph{\textbf{SERA Enhanced Information Processing Efficiency and Decision Confidence}}

\begin{figure}[ht]
    \centering
    \includegraphics[width=\linewidth]{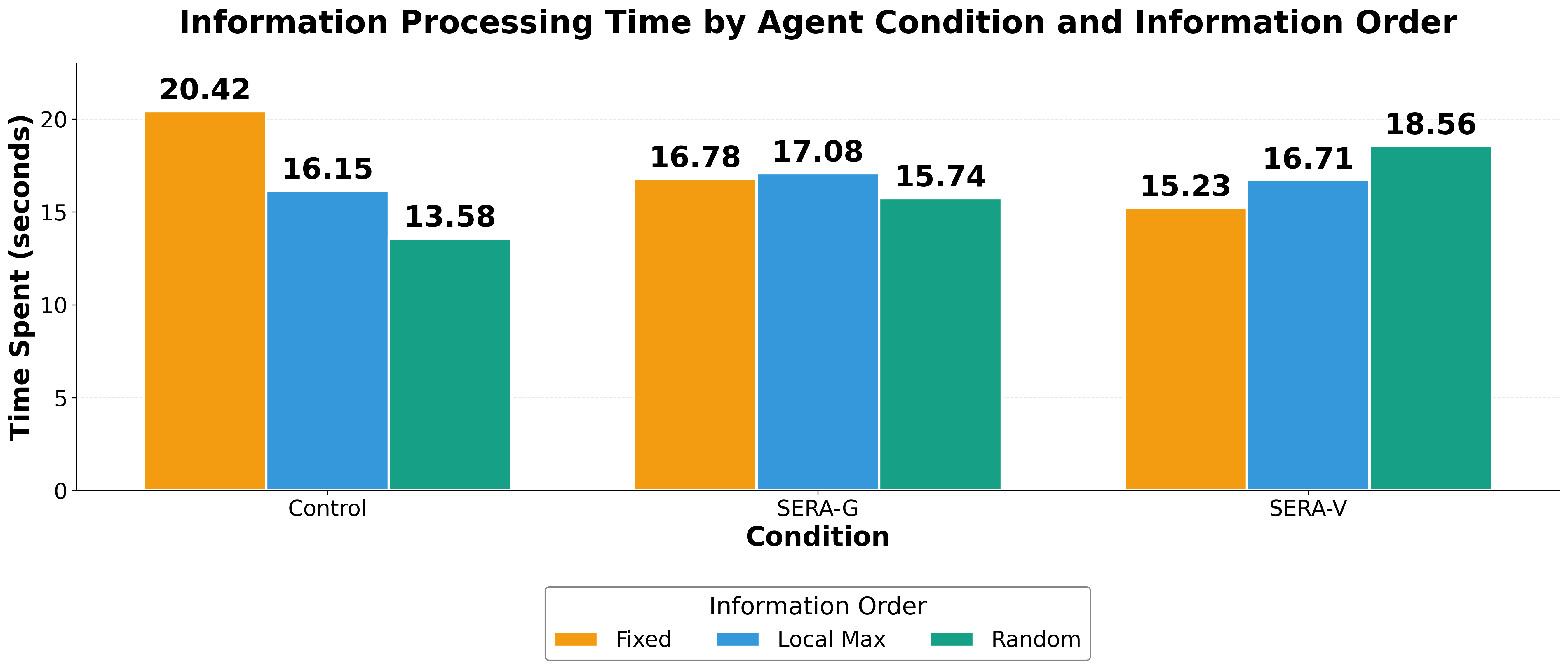}
    \caption{Processing efficiency (time per information unit) by agent condition and environmental uncertainty.}
    \label{fig:processingeff}
\end{figure}

We operationalized processing efficiency as the amount of time spent per unit of information. We examined the fixed effects of agents and environmental uncertainty on processing efficiency, considering the random effects of participants and scenarios (see Model 2 in Table~\ref{tab:processingeff}). We found significant interaction effects of agents and environmental uncertainty, showing that people spent much shorter time processing each information sampled in the control than SERA conditions when information was presented in random distribution ($B=-5.49$, $SE=2.28$, $t=-2.41$, $p<.05$), and much longer time processing each information in the SERA conditions than the control one when the information was presented in decremental distribution (see Figure~\ref{fig:processingeff}).

\begin{table}[ht]
\centering
\caption{Mixed-effects model predicting processing efficiency by agent condition and environmental uncertainty (Model 2).}
\label{tab:processingeff}
\begin{tabular}{lccc}
\toprule
\textbf{Predictor} & \textbf{B (SE)} & \textbf{t} &  \\
\midrule
Intercept & 16.63 (1.76) & 9.43** \\
Agent 1. Control vs SERA & −0.01 (1.98) & −0.01 \\
Agent 2. SERA-G vs SERA-V & −0.38 (2.28) & −0.17 \\
Environmental Uncertainty 1. Non-Random vs Random & 1.04 (1.07) & 0.98 \\
Environmental Uncertainty 2. Decremental vs Local-Opt & −0.98 (1.22) & −0.81 \\
Agent 1 × Environmental Uncertainty 1 & −5.49 (2.28) & −2.41* \\
Agent 2 × Environmental Uncertainty 1 & 3.90 (2.63) & 1.49 \\
Agent 1 × Environmental Uncertainty 2 & 5.27 (2.60) & 2.03* \\
Agent 2 × Environmental Uncertainty 2 & −0.91 (3.00) & −0.30 \\
\midrule
Log Likelihood & \multicolumn{2}{c}{−553.6} &  \\
\bottomrule
\multicolumn{3}{l}{\textit{Note.} *$p < .05$; **$p < .01$}
\end{tabular}
\end{table}

Further, after making the decision, participants rated the perceived confidence of the information sampled in the scenario. We examined the effects of agents and environmental uncertainty on perceived confidence of information, with random effects of participants and scenarios. Findings showed significant main effects of agent conditions, which participants rated higher perceived confidence of information in SERA-G than SERA-V ($B=0.74$, $SE=0.30$, $t=2.42$, $p<.05$), suggesting the gist format would be a preferred format to help adults feel more confident about the information.

\subsubsection{RQ2: Effects of SERA Type and Environmental Uncertainty on Decision-making Behavior} 

\begin{figure}[ht]
    \centering
    \includegraphics[width=\linewidth]{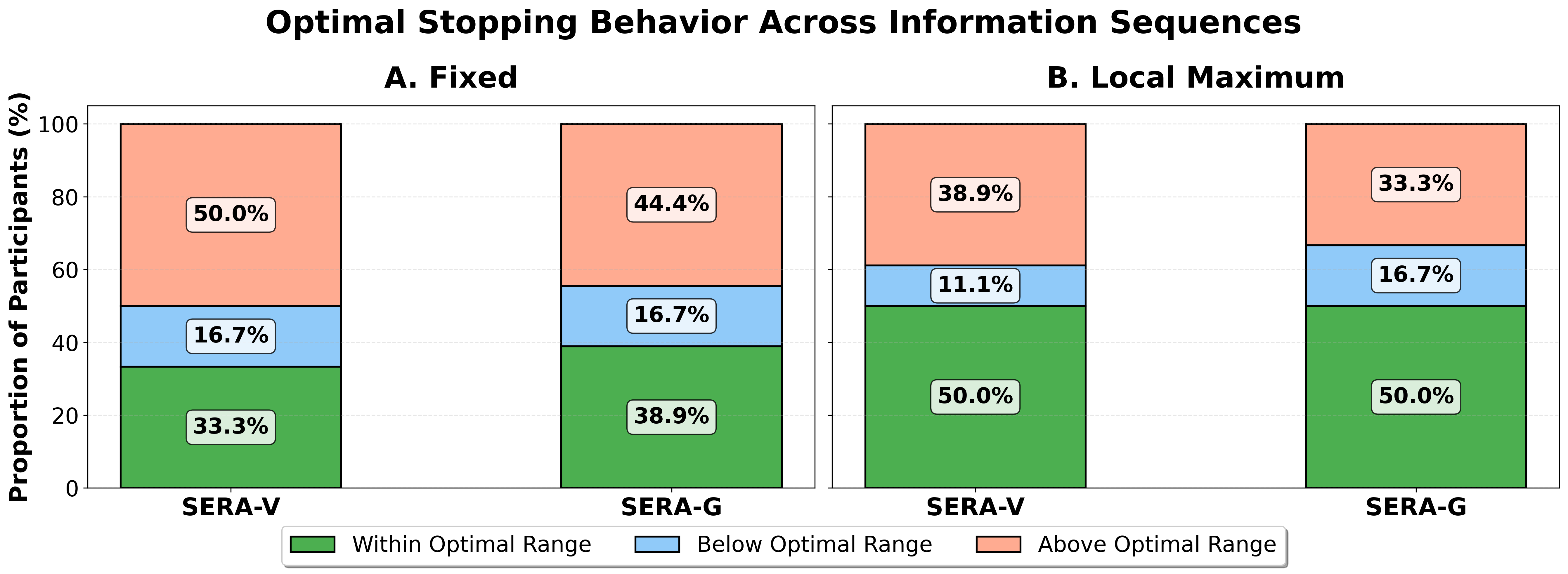}
    \caption{Main Study: Distribution of stopping behavior relative to optimal information range by SERA feedback type and environmental uncertainty.}
    \label{fig:main_optimalstop}
\end{figure}

\paragraph{\textbf{SERA-G was Associated with Less Oversampling Under Randomized Sequences}}
The main study further examined how SERA feedback type and environmental uncertainty influenced optimal stopping behavior using chi-square tests (see~\ref{appendix:optimalrange} for range definitions). Figure~\ref{fig:main_optimalstop} illustrates the distribution of stopping patterns across experimental conditions.

\begin{itemize}
    \item \textbf{Effect of SERA feedback type.} Comparing SERA-G and SERA-V, gist-based feedback was associated with lower oversampling, particularly under conditions of higher uncertainty. Across all conditions, 44.4\% of SERA-G participants stopped within the optimal range compared to 41.7\% for SERA-V, $\chi^{2}(2) = 0.26$, $p = .880$, $V = 0.06$. When examined by environmental uncertainty, both feedback types showed comparable calibration in the Decremental distribution (SERA-G = 38.9\%, SERA-V = 33.3\%; $\chi^{2}(2) = 0.14$, $p = .934$, $V = 0.06$). In the Local Optimum (Local-Opt) distribution, both achieved 50.0\% optimal stopping ($\chi^{2}(2) = 0.28$, $p = .871$, $V = 0.09$). These differences did not reach statistical significance; however, the overall pattern is consistent with gist-oriented feedback supporting more efficient search regulation in these complex scenarios.

    \item \textbf{Effect of environmental uncertainty.} Comparing Decremental and Local-Opt distributions revealed a non-significant trend favoring the Local-Opt condition, $\chi^{2}(2) = 1.43$, $p = .489$, $V = 0.14$. Half of participants (50.0\%) stopped within the optimal range in Local-Opt compared to 36.1\% in Decremental distribution. Oversampling (\textit{Slower} category) occurred more frequently in the Decremental condition (47.2\%) than in Local-Opt (36.1\%), suggesting that presenting locally informative items earlier supported more adaptive stopping decisions by accelerating evidence accumulation.
\end{itemize}

\paragraph{\textbf{SERA Increased Early-Stage Switching and Overall Exploration}}
Exploration was operationalized as the number of switches between options and total information sampled. Mixed-effects analyses revealed a significant main effect of agent condition, with participants making more switches in SERA conditions than control ($B = 5.58$, $SE = 2.37$, $t = 2.36$, $p < .05$).

Participants in SERA conditions exhibited significantly more switches during the first third of the decision period compared to control ($B = 2.31$, $SE = 0.76$, $t = 3.06$, $p < .01$), indicating SERA feedback encouraged early-stage exploration and comparison between alternatives. Participants in SERA conditions also sampled substantially more information overall than control ($B = 9.38$, $SE = 2.65$, $t = 3.54$, $p < .001$). No significant main or interaction effects of environmental uncertainty were observed. These findings extend the preliminary results, demonstrating that SERA feedback promotes more active exploration and information gathering, particularly early in the decision process.

\paragraph{\textbf{Distinct Search Strategies Emerged Across SERA Feedback Types}}

To examine differences in information search strategy, the main study expanded the classification scheme to include an additional \textit{Interchangeable} category (frequent, unstructured switching). Two independent coders classified early-phase search patterns with substantial agreement (Cohen’s $\kappa = 0.83$).

A chi-square test revealed no significant association between decision scenario and search pattern, $\chi^{2}(4, N = 162) = 2.04$, $p = .729$, $V = .08$, indicating consistent strategy use across contexts. However, the majority of participants adopted \textit{Piecewise} sampling across all scenarios (67–74\%), suggesting a dominant sequential evaluation tendency.  

Search patterns differed significantly by SERA feedback type, $\chi^{2}(4, N = 162) = 15.74$, $p = .003$, $V = .22$ (Table~\ref{tab:sample_feedback_main}). \textit{SERA-V} participants were more likely to adopt \textit{Comprehensive} strategies (31\%) compared to No-SERA (19\%), while \textit{SERA-G} participants more often engaged in \textit{Interchangeable} exploration (20\%) relative to No-SERA (7\%). Post hoc pairwise comparisons revealed a significant difference between SERA-V and SERA-G, $\chi^{2}(2) = 14.89$, $p < .001$, $V = .37$, with SERA-V promoting systematic comparison and SERA-G supporting flexible switching. 

\begin{table}[ht]
\centering
\caption{Distribution of early-phase search patterns (Piecewise, Comprehensive, Interchangeable) by SERA feedback type.}
\label{tab:sample_feedback_main}
\begin{tabular}{lcccc}
\toprule
\textbf{Search Pattern} & \textbf{No-SERA} & \textbf{SERA-G} & \textbf{SERA-V} & \textbf{Total} \\
\midrule
Piecewise        & 40 (74\%) & 37 (69\%) & 36 (67\%) & 113 \\
Comprehensive    & 10 (19\%) & 6 (11\%)  & 17 (31\%) & 33 \\
Interchangeable  & 4 (7\%)   & 11 (20\%) & 1 (2\%)   & 16 \\
\bottomrule
\end{tabular}
\end{table}

\begin{figure}[h]
    \centering
    \includegraphics[width=\linewidth]{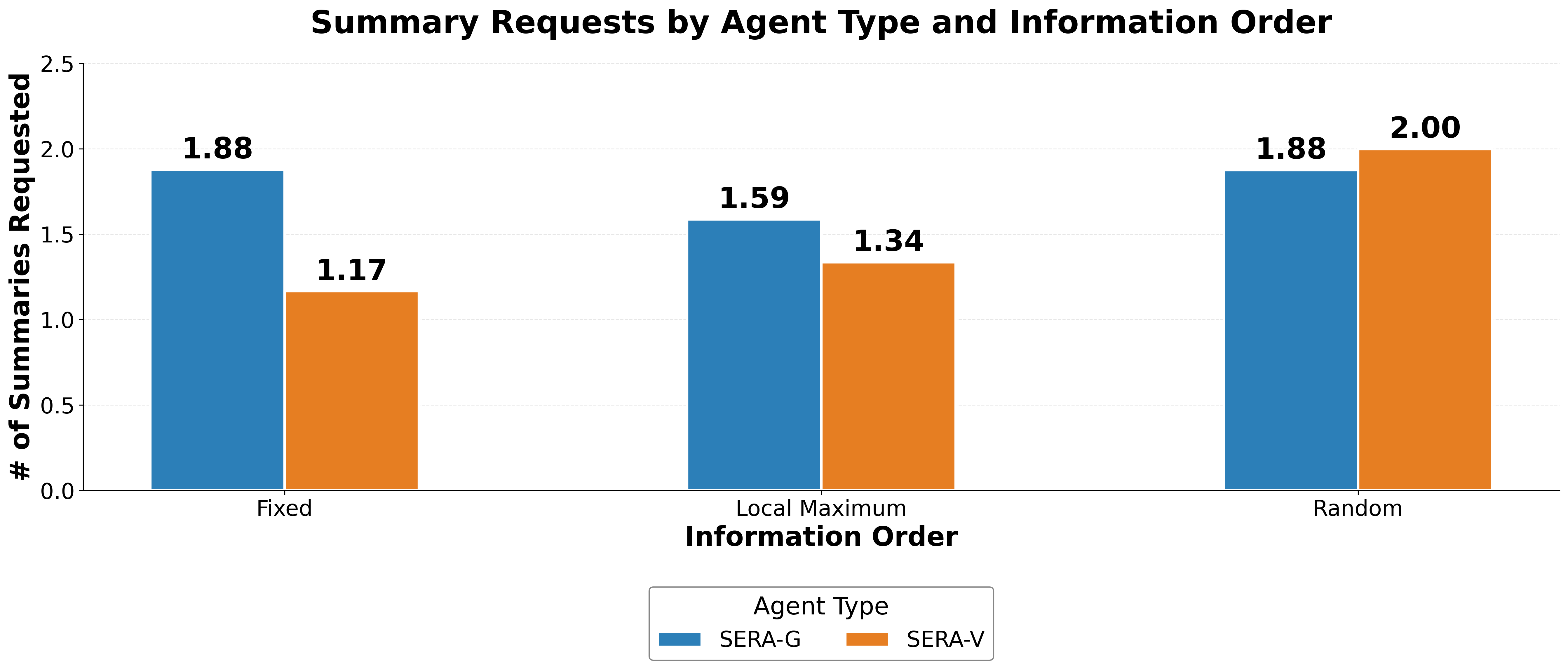}
    \caption{Number of summaries requested by SERA feedback type and environmental uncertainty.}
    \label{fig:summaryrequested}
\end{figure}

\paragraph{\textbf{Participants Engaged More with SERA-G than SERA-V, particularly in Structured Sequences}}
We analyzed the number of summaries requested during SERA conditions as a measurement of engagement. Mixed-effects models examined the fixed effects of SERA agent type (gist vs. verbatim) and environmental uncertainty on summaries requested per scenario, with random effects of participants and scenarios (Model 3, Table~\ref{tab:summaryrequests}). Results showed a significant interaction between agent type and environmental uncertainty ($B = -0.61$, $SE = 0.31$, $t = -1.96$, $p < .05$), indicating participants requested more summaries in SERA-G than SERA-V particularly in structured sequences (Figure~\ref{fig:summaryrequested}). Compared to SERA-V, SERA-G summaries were requested more frequently when information was presented in Decremental and Local-Opt distributions, suggesting participants utilized gist summaries more to integrate sampled information and make decisions. However, both SERA conditions generated similar summary request rates when information was presented randomly.

\begin{table}[ht]
\centering
\caption{Mixed-effects model predicting summaries requested by agent type and environmental uncertainty (Model 3).}
\label{tab:summaryrequests}
\begin{tabular}{lcc}
\toprule
\textbf{Predictor} & \textbf{B (SE)} & \textbf{t} \\
\midrule
Intercept & 1.64 (0.18) & 9.10** \\
Agent Type. SERA-G vs SERA-V & $-$0.28 (0.36) & $-$0.79 \\
Environmental Uncertainty 1. Non-Random vs Random & $-$0.45 (0.16) & $-$2.91** \\
Environmental Uncertainty 2. Decremental vs Local-Opt & $-$0.06 (0.18) & $-$0.36 \\
Agent Type $\times$ Environmental Uncertainty 1 & $-$0.61 (0.31) & $-$1.96* \\
Agent Type $\times$ Environmental Uncertainty 2 & 0.46 (0.36) & 1.30 \\
\midrule
Log Likelihood & \multicolumn{2}{c}{$-$149.7} \\
\bottomrule
\multicolumn{3}{l}{\textit{Note.} *$p < .05$; **$p < .01$}
\end{tabular}
\end{table}

\subsubsection{RQ3: Subjective Experience and Perceived Usefulness of SERA}

\begin{figure}[ht]
    \centering
    \includegraphics[width=1\linewidth]{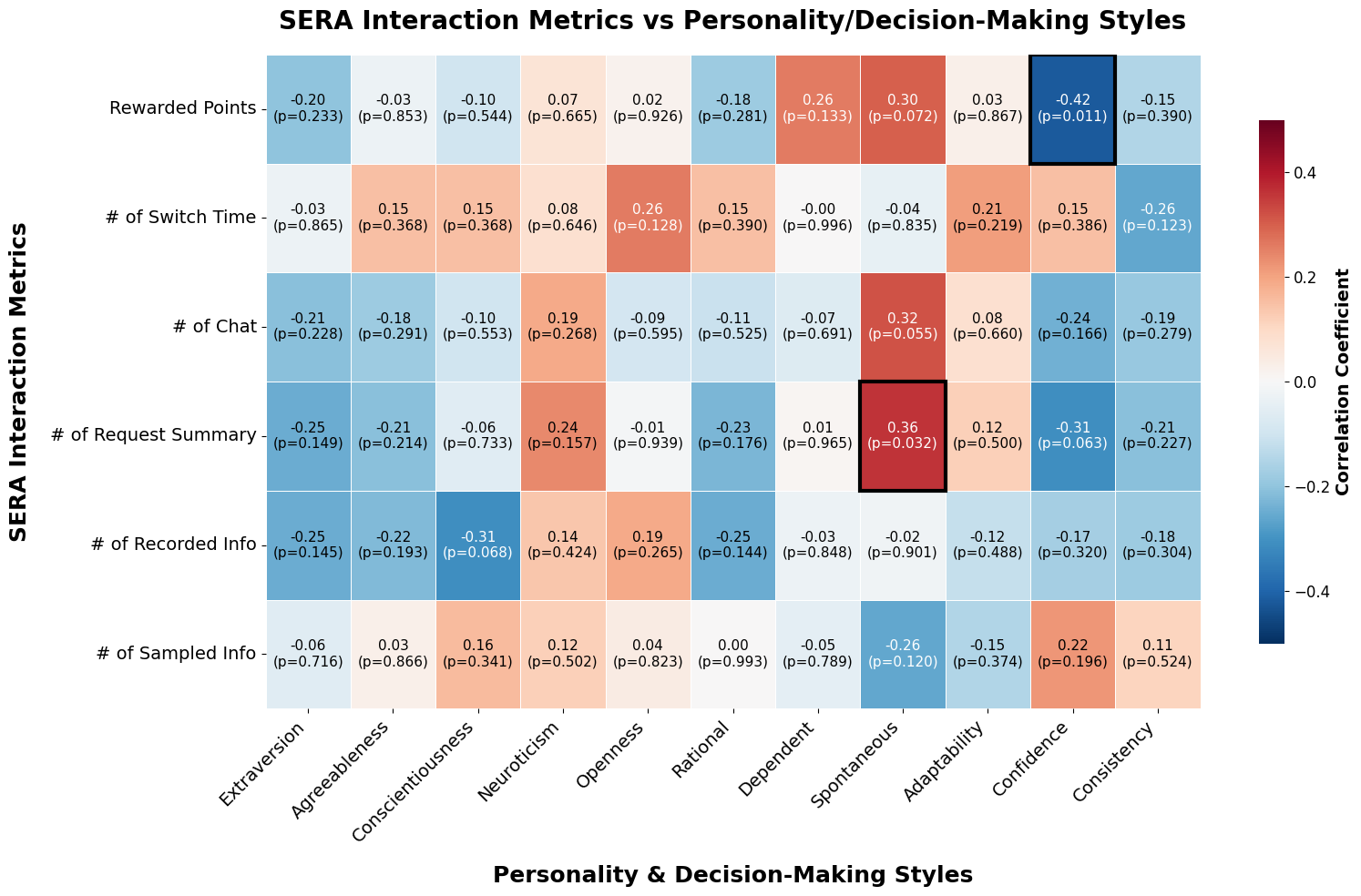}
    \caption{Correlations between SERA interaction metrics and individual differences in personality and decision-making style.}
    \label{fig:individual_diff}
\end{figure}

\paragraph{\textbf{Individual Differences and SERA Interaction Patterns}}
We examined how participants’ personality traits and decision-making styles related to their behavioral engagement with SERA (Figure~\ref{fig:individual_diff}). Correlational analyses showed that individuals with a more \textit{rational} decision-making style interacted more actively with the system, requesting summaries more frequently ($r = .36$, $p = .032$) and showing a marginal trend toward initiating more chats ($r = .32$, $p = .055$). In contrast, higher \textit{confidence} scores were negatively associated with total rewarded points ($r = -.42$, $p = .011$), suggesting that more confident participants relied less on SERA feedback. No other correlations reached significance.

\begin{figure}[ht]
    \centering
    \includegraphics[width=1\linewidth]{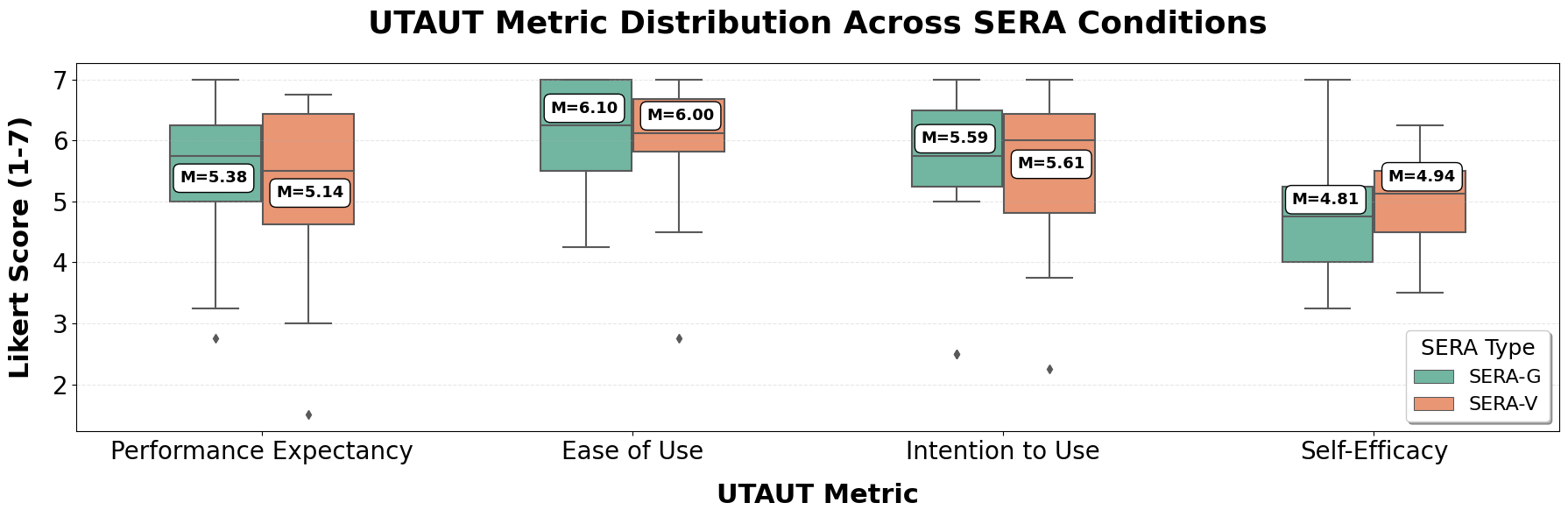}
    \caption{Self-reported UTAUT ratings (Performance Expectancy, Effort Expectancy, Intention to Use, Self-Efficacy) by SERA condition.}
    \label{fig:utaut}
\end{figure}

\paragraph{\textbf{Technology Acceptance (UTAUT)}}
A second mixed-effects model examined UTAUT ratings across four subscales. No significant main effect of SERA condition was found ($\beta = -0.10$, $p = .794$). Descriptively, SERA-G participants rated Performance Expectancy ($M_{\text{SERA-G}} = 5.38$, $M_{\text{SERA-V}} = 5.14$, $d = 0.18$) and Ease of Use ($M_{\text{SERA-G}} = 6.10$, $M_{\text{SERA-V}} = 6.00$, $d = 0.11$) slightly higher than SERA-V participants (Figure~\ref{fig:utaut}). Both conditions received favorable usability and acceptance ratings, showing participants perceived SERA as supportive and easy to use in decision-making tasks.

\subsection{Main Study Discussion}
The main study demonstrates that SERA feedback improved decision accuracy, confidence, and processing efficiency, with the strongest benefits in high-uncertainty environments where information value was unpredictable. This interaction between tool and environment is central to our contribution: SERA provided the greatest advantage precisely when users could not rely on environmental structure to guide their stopping decisions. Gist-based feedback was associated with more adaptive search patterns and lower oversampling, whereas verbatim feedback promoted systematic but slower analytic comparison. The distinct search strategies that emerged across feedback types---flexible switching under gist conditions versus systematic comparison under verbatim conditions---suggest that feedback representation shapes not only when users stop but how they search. Participants across both SERA conditions reported positive decision-making experiences and high usability, indicating that the observed behavioral differences reflect genuine strategy adaptation rather than differential engagement.

\section{General Discussion}
\subsection{Summary of Results}
This work examines how AI feedback representations interact with environmental uncertainty to shape decision-making with information search. Across two controlled studies, SERA improved decision accuracy and confidence relative to no feedback, with the critical finding that these benefits depended on environmental conditions: SERA provided the largest advantages in high-uncertainty environment where perceived information gain was unpredictable over time. Beyond decision outcomes, feedback representation systematically influenced decision processes during information search under uncertainty. Gist-based feedback was associated with earlier convergence and less oversampling, consistent with meaning-based integration and sufficiency-oriented stopping, whereas verbatim feedback promoted extended comparison and continued exploration through detail-focused verification. The interaction between feedback type and environmental uncertainty indicates that effective decision support requires attention to both tool design and the structure of the environment in which the tool is deployed.

\subsection{Extending Fuzzy-Trace Theory to Human--AI Decision Support}

Our study extends Fuzzy-Trace Theory (FTT) to human--AI decision support by demonstrating that gist and verbatim can be operationalized as alternative feedback representations that regulate decision processes under uncertainty. FTT characterizes gist and verbatim as internal cognitive representations that support memory and judgment \citep{reyna2008theory,reyna2015,brainerd2002}. Our results indicate that these representations can be instantiated in AI feedback at the interface level, shaping how decision makers allocate attention, integrate evidence, and determine when additional search is no longer worthwhile.

Across studies, gist feedback may help participants regulate the tradeoff between exploration and exploitation, and support more optimal stopping when further search offered limited marginal values in information gain. Verbatim feedback  encouraged continued verification and led to more exploration when evidence remained ambiguous. These patterns suggest that feedback representation can shape cognitive effort allocation and inferences about the information environment during decision-making under uncertainty, beyond merely describing internal cognitive states.
This perspective aligns with recent HCI work showing that generative systems can leverage gist-oriented representations to support interpretability and user reasoning \citep{gleaves2024impact}, and that manipulating representational form can scaffold cognition in complex tasks such as writing and sensemaking \citep{lin2024rambler}. Unlike prior research that primarily frames human--AI decision support in terms of transparency, usability, or trust \citep{lai2023empirical,wang2023human,raees2024explainable}, our contribution is to demonstrate that aligning AI feedback representations with cognitive representations provides a principled design lever for shaping how people organize, regulate, and terminate information search in uncertain information environments.

\subsection{Cognitive Regulation in Search under Uncertainty}
Beyond the theoretical integration, our results revealed that feedback representations shaped how users managed cognitive effort during decision-making. In the gist condition, participants often stopped when “the summary made sense,” they “felt enough,” or noticed a “clear difference” between options. Such cues indicate a meaning-based stopping rule guided by perceived understanding of the information environment. Gist summaries helped users identify when additional search provided little new information, supporting efficient self-regulation and adaptive effort allocation \citep{payne1993adaptive, Lieder_Griffiths_2020}.
Participants in the verbatim condition showed the opposite pattern. They reported continuing “until I read everything” or “checked all points,” reflecting a compensatory and information-intensive rule. The literal feedback encouraged verification and detailed checking, which increased cognitive load and prolonged monitoring, similar to experience-based strategies where confidence depends on accumulated evidence \citep{hertwig2009}.
Supplemented by our main study results in participants' optimal search behavior, these findings suggest that feedback representation can steer users toward distinct regulatory modes.  Future work should examine how dynamically adjusting feedback representation helps decision makers balance efficiency and accuracy in complex tasks.

\subsection{Design Implications}

\subsubsection{Design Implication 1: Adaptive Feedback Representation for Decisions under Uncertainty}
Our findings suggest that decision-support systems can operationalize Fuzzy Trace Theory by adaptively shifting between gist and verbatim feedback to fit the demands of decision-making under uncertainty. Across three levels of environmental uncertainty, SERA's gist feedback improved efficiency and integration when predictability was low, while more detailed feedback supported analytic comparison when information followed a clearer pattern. This pattern indicates that systems need to vary the granularity of feedback to match uncertainty gradients in the environment and to balance cognitive load with precision. Prior research on adaptive transparency and feedback control shows that dynamically adjusting information depth improves trust calibration and task performance \citep{akash2020adaptive,gomez2025taxonomy}. Work on information granularity similarly finds that feedback detail shapes comprehension and decision speed \citep{Wilke2016GranularCA,kagrecha2025granular}. These results motivate designs that modulate feedback representation in real time as uncertainty shifts, maintaining coherence when structure is weak and supporting detailed analysis when structure is strong.

\subsubsection{Design Implication 2: Making Environmental Structure and Uncertainty Visible}
Our findings indicate that effective decision support in search under uncertainty tasks should help users reason about the information environment, not only about individual options. Participants needed to monitor the information gain across the course of search, and SERA’s feedback representation supported this implicit assessment by summarizing trends across sampled information. Future systems can make such environmental structure more explicit, for example by indicating how consistent current evidence is over time, whether newly sampled information continues prior trends, or how much additional search is likely to change the relative standing of options. Providing these higher level cues about predictability and remaining uncertainty would help decision makers calibrate expectations about the environment, set more appropriate stopping rules, and allocate cognitive effort where further search is most likely to be informative.

\subsubsection{Design Implication 3: Supporting Search and Stopping as Core Interaction Goals}
We found that SERA influenced how users gathered and evaluated information rather than only what choices they made. Gist feedback encouraged participants to stop when the information felt sufficient, while verbatim feedback promoted more exhaustive verification when evidence was structured. These findings suggest that decision-support systems should not focus solely on generating recommendations but also guide the reasoning process itself. Recent studies have made similar observations. \citet{zhang2024forward} showed that forward AI support guiding pilots’ information search and stopping improved decision quality compared to traditional recommendation systems. \citet{gomez2025taxonomy} reviewed AI-assisted decision-making research and called for interactive systems that structure and monitor users’ information search. Likewise, \citet{van2025human} emphasized that explainable AI in healthcare should align with clinicians’ cognitive workflows, supporting how they think rather than what they decide. Based on these insights, future decision-support tools should include mechanisms that observe users’ behavioral signals, such as search pace or hesitation, and offer timely cues to help reflection and stopping. Embedding such process-level regulation would shift AI systems from static output generators to active partners that support users’ agency and adaptability during complex decisions.

\section{Limitations}
There are several limitations to this study. First, although SERA demonstrated promising effects on decision accuracy and regulation, the tasks were conducted in short, text-based scenarios that simplified real-world decision complexity. Genuine decisions in medical, financial, or social contexts are dynamic and emotionally charged, often involving multiple stakeholders, which limits the ecological validity of the current findings. Second, the participant samples were relatively small and homogeneous, consisting mostly of young, educated, English-speaking individuals recruited online or from a university setting. This demographic bias may overestimate usability and alignment with LLM-based feedback, suggesting a need for broader and cross-cultural replication. Third, participants interacted with SERA in a single session with only three short scenarios, preventing observation of longer-term adaptation, trust development, or fatigue. Future studies should examine repeated or longitudinal use to capture how users co-adapt with AI support over time. Another limitation concerns the underlying language mode, while the GPT-based prototype proved effective for most participants, there are still limitations in the accuracy and efficiency of the models. Although SERA’s feedback successfully conveyed both gist and verbatim representations of information, the processing time is an ongoing problem that causes lags or interruptions when making decisions. Future research should focus on refining these aspects of SERA’s design to enhance its reliability and utility in complex decision-making environments. Finally, this study mainly served as a proof of concept rather than a fully adaptive system, as feedback representation levels were fixed rather than dynamically adjusted. Future work should extend SERA to higher-stakes domains and explore real-time adaptation that personalizes feedback based on users’ cognitive and contextual states.

\section{Conclusion}
Overall, our study demonstrated that SERA improved decision accuracy, confidence, and processing efficiency in information search under uncertainty, with the strongest benefits when environmental uncertainty made the environment less predictable. Gist-based feedback was associated with more regulated effort and lower oversampling under higher uncertainty, whereas verbatim feedback encouraged more systematic but extended exploration when informational structure was clearer. Participants also reported positive experiences and high usability across both feedback types, with engagement patterns varying by individual decision style and confidence. These findings highlight that aligning AI feedback representation with how people regulate effort in uncertain environments can enhance self regulation and decision quality. SERA provides a proof of concept for large language model based decision support that scaffolds users’ reasoning about both options and the surrounding information environment, and can inform future adaptive and context-aware systems for complex, information-rich settings.

\section*{Acknowledgements}
We thank all participants who took part in our study.

\section*{Declaration of Interest}
The authors declare no conflicts of interest related to this research.

\section*{Funding}
This research was not funded.

\section*{CRediT Author Statement}
\textbf{Kexin Quan:} Conceptualization, Methodology, Software, Data curation, Formal analysis, Writing - original draft, Writing - review \& editing. \textbf{Jessie Chin:} Supervision, Methodology, Formal analysis, Writing - review \& editing.

\section*{Declaration of generative AI and AI-assisted technologies in the manuscript preparation process}

During the preparation of this work the author(s) used ChatGPT in order to refine wording, correct grammar and spelling, and assist with literature searches. After using this tool, the author(s) reviewed and edited the content as needed and take(s) full responsibility for the content of the published article.

\bibliographystyle{elsarticle-harv}
\bibliography{references}

\appendix
\label{sec: appendix}

\section{Questionnaires}
\subsection{Checkpoint Survey}
\label{appendix: ckpt_survey}
How much would you agree with the following statement? (Likert Scale, 1-7)

•	I am confident about the decision I made. 

•	I find the information easy to understand. 

•	I am able to consider all the different information I learned to make the decision. 

\subsection{Pre-Survey Questions by Construct}
\label{appendix: pre-survey}
\begin{table}[H]
\centering
\caption{BFI-10 Questionnaire (Likert Scale, 1–5). Certain items are reverse-scored (R).}
\vspace{0.5cm}
\begin{tabular}{|p{3.1cm}|p{9cm}|}
\hline
\textbf{Trait} & \textbf{Survey Questions} \\
\hline
Extraversion &
1R. Is reserved \\
& 6. Is outgoing, sociable \\
\hline
Agreeableness &
2. Is generally trusting \\
& 7R. Tends to find fault with others \\
\hline
Conscientiousness &
3R. Tends to be lazy \\
& 8. Does a thorough job \\
\hline
Neuroticism &
4R. Is relaxed, handles stress well \\
& 9. Gets nervous easily \\
\hline
Openness to Experience &
5R. Has few artistic interests \\
& 10. Has an active imagination \\
\hline
\end{tabular}
\label{tab: bfi-10}
\end{table}

\begin{table}[H]
\centering
\caption{General Decision-making Style (GDMS) Questionnaire, Grouped by Decision-making Style (Likert Scale, 1–5)}
\vspace{0.5cm}
\begin{tabular}{|p{2.1cm}|p{12.5cm}|}
\hline
\textbf{Style} & \textbf{Survey Questions} \\
\hline
Rational &
1. I double-check my information sources to be sure I have the right facts and analysis. \\
& 2. I make decisions in a logical and systematic way. \\
& 3. My decision-making requires careful thought. \\
& 4. When making a decision, I consider various options in terms of a specific goal. \\
& 5. I explore all of my options before making a decision. \\
\hline
Intuitive &
6. When I make decisions, I rely upon my instincts. \\
& 7. When I make decisions, I tend to rely on my intuition. \\
& 8. I generally make decisions that feel right to me. \\
& 9. When I make a decision, it is more important for me to feel the decision is right than to have a rational reason for it. \\
& 10. When making important decisions, I trust my inner feelings and reactions. \\
\hline
Dependent &
11. I often need the assistance of other people when making important decisions. \\
& 12. I rarely make important decisions without consulting other people. \\
& 13. If I have the support of others, it is easier for me to make important decisions. \\
& 14. I use the advice of other people in making my important decisions. \\
& 15. I like to have someone to steer me in the right direction when I am faced with important decisions. \\
\hline
Avoidant &
16. I avoid making important decisions until the pressure is on. \\
& 17. I postpone decision-making whenever possible. \\
& 18. I often procrastinate when it comes to making important decisions. \\
& 19. I generally make important decisions at the last minute. \\
& 20. I put off making many decisions because thinking about them makes me uneasy. \\
\hline
Spontaneous &
21. I generally make snap decisions. \\
& 22. I often make decisions on the spur of the moment. \\
& 23. I make quick decisions. \\
& 24. I often make ``impulsive'' decisions. \\
& 25. When making decisions, I do what seems natural at the moment. \\
\hline
\end{tabular}
\end{table}

\subsection{Post-Survey Questions by Construct}
\label{appendix:post-survey}
\begin{table}[H]
\centering
\caption{Adapted UTAUT Items (Likert Scale, 1-7)}
\vspace{0.5cm}
\begin{tabular}{|p{2.2cm}|p{12cm}|}
\hline
\textbf{Category} & \textbf{Post-Survey Questions} \\
\hline
Performance Expectancy & 
1. I would find the system useful in making decisions

2. Using the system enables me to accomplish tasks more quickly

3. Using the system increases my productivity

4. I will increase the quality of the output of making decisions \\
\hline
Effort Expectancy & 
5. My interaction with the system would be clear and understandable

6. It would be easy for me to become skillful at using the system

7. I would find the system easy to use

8. Learning to operate the system is easy for me \\
\hline
Intention to Use & 
9. Using the system is a good idea

10. The system makes making decisions more interesting

11. Working with the system is fun

12. I like working with the system \\
\hline
Self-Efficacy & 
13. If there was no one around to tell me what to do as I go

14. If I could call someone for help if I got stuck

15. If I had a lot of time to complete the job for which the software was provided

16. If I had just the built-in help facility for assistance \\
\hline
\end{tabular}
\label{tab: utaut}
\end{table}

\section{Summarization using SERA}
\subsection{Summary Prompts}
We developed custom prompts for SERA's summarization feature, which are listed below.

\label{appendix:summary_prompts}

\begin{promptbox}[SERA-Gist Prompt]
\textbf{Role:} You are a professional AI chatbot tasked with providing concise gist statements for non-expert decision-makers. Your role is to analyze and synthesize the key differences between two options to facilitate informed decision-making. For each pair of options, extract and compare their key aspects and present the conclusion in a succinct gist statement. If the input is empty, output ``You haven't recorded any important information yet!''

\vspace{0.3cm}
\textbf{Example:}

\begin{examplebox}
\textit{Input:} A: The monthly rent amount for ApartmentA is \$1350, with an extra fee of \$150/mon for amenities including gym access and high-speed internet., B: The monthly rent for ApartmentB is \$1050, with extra fees of \$250/mon that include utilities and parking., A: ApartmentA's proximity to school allows for easy access to libraries, study groups, and campus events, reducing the need for transportation., B: ApartmentB units are spacious with a balcony or patio.

\vspace{0.2cm}
\textit{Output:} For decision-makers valuing affordability, amenities like gym access and high-speed internet, and proximity to school reducing transportation needs, ApartmentA is a favorable choice. Conversely, ApartmentB offers spacious units with personal outdoor spaces with inclusive utilities and parking for those seeking comfort and outdoor living options.
\end{examplebox}
\end{promptbox}

\vspace{0.5cm}

\begin{promptbox}[SERA-Verbatim Prompt]
\textbf{Role:} You are a smart and professional AI chatbot tasked with providing concise verbatim statements for decision-makers who are not experts in the field. Your role is to analyze and synthesize the key differences between two options, facilitating quick and informed decision-making. For each pair of options, you will extract and compare their essential aspects and present the conclusion in a succinct verbatim statement that is analyzed in different criteria. If the input is empty, simply output ``You haven't recorded any important information yet!''

\vspace{0.3cm}
\textbf{Example:}

\begin{exampleboxpurple}
\textit{Input:} A: The monthly rent amount for Greenwood Apartments is \$1350, with an additional fee of \$150/month for amenities including gym access and high-speed internet., B: The monthly rent amount for Lakeside Residences is \$1050, along with additional fees of \$250 per month that include utilities and parking., A: Greenwood Apartments's proximity to the school allows for easy access to libraries, study groups, and campus events, reducing the need for transportation., B: The apartment units are spacious with a balcony or patio, providing a personal outdoor space for residents.

\vspace{0.2cm}
\textit{Output:} Greenwood Apartments: \$1350/month rent + \$150/month amenities (gym, internet); close to school for easy campus access. Lakeside Residences: \$1050/month rent + \$250/month utilities and parking; spacious units with balconies or patios.
\end{exampleboxpurple}
\end{promptbox}

Below provides the structure of one participant's recorded information list, along with the generated gist/verbatim summary output from SERA: 
\begin{figure}[H]
    \centering
    \includegraphics[width=1.2\linewidth]{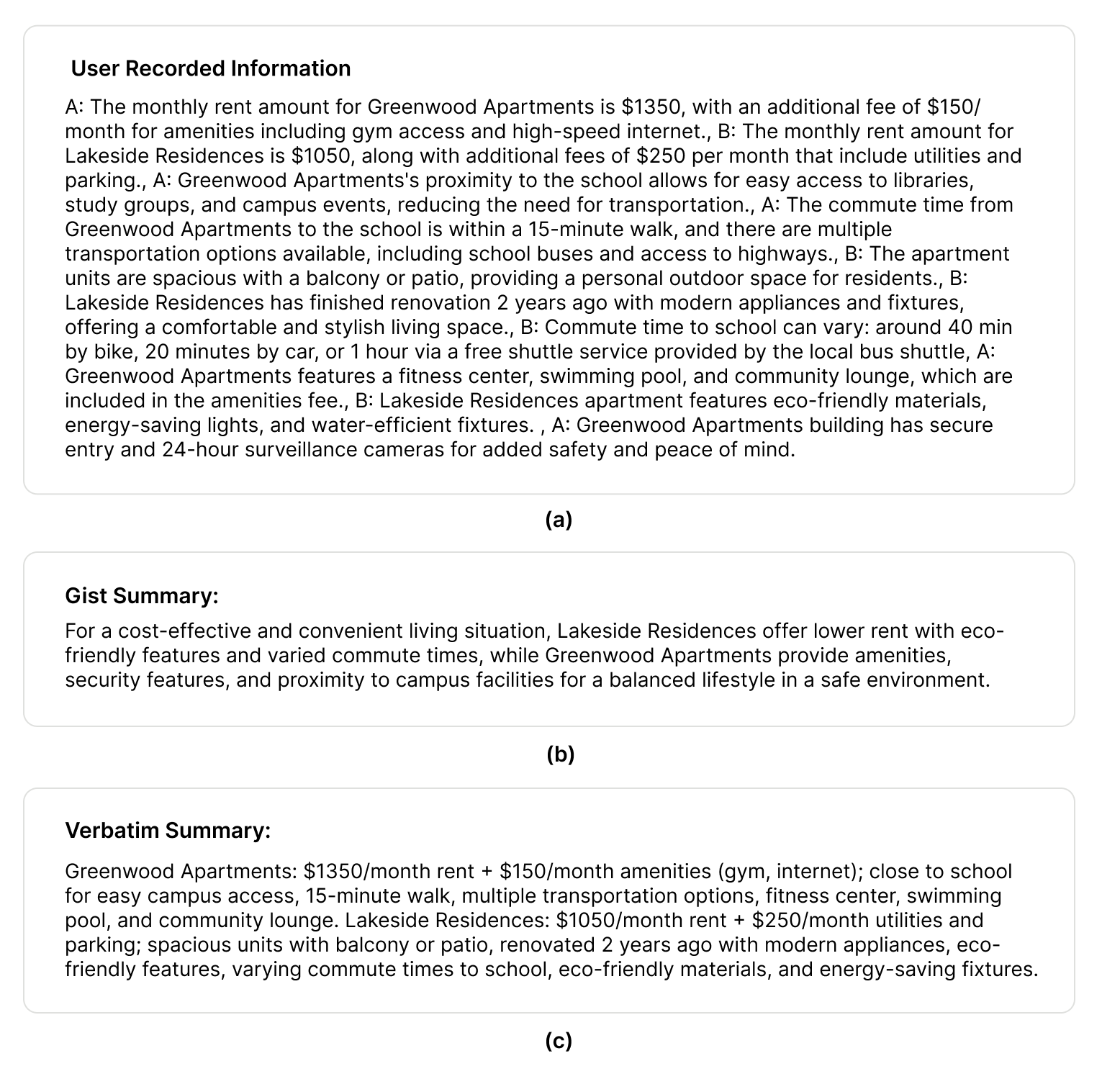}
    \caption{Example of SERA’s Gist vs. Verbatim Summary Generation Based on User Input. (a) User's recorded information for options A and B. (b) Gist and (c) Verbatim summaries generated by SERA.}
    \label{fig:sera_example}
\end{figure}

\subsection{Decision-making Scenarios}
The two sections below provide detailed methods on how information lists are formed in two studies.
\label{appendix:dm_scenarios}

\subsubsection{Designing Information Pieces for Preliminary Study}

\label{appendix:prelim_designinfo}
Recognizing the necessity for a consistent depth and breadth of information, we identified 10 key attributes that individuals commonly consider when making decisions in each scenario. These factors were chosen based on prior research and their relevance to the scenarios at hand. Each attribute was considered to ensure comprehensive coverage of the decision-making landscape. For instance, when designing information for Stock Investment (Scenario 2), we included features such as the P/E ratio, dividends and yield for two stocks' financial performance and investments. Based on 10 attributes of each scenario, we created both options’ information messages. All information pieces were matched in length (average 15–19 words), with no statistical differences between options across scenarios (Scenario 1: $p=0.704$; Scenario 2: $p=0.216$; Scenario 3: $p=1.0$).   

\begin{longtable}{|p{5cm}|p{9cm}|}
\caption{Preliminary Study Decision-making (DM) Scenarios} \label{tab:study1_personas} \\
\hline
\textbf{Scenario} & \textbf{DM Scenario Description} \\
\hline
\endfirsthead

\hline
\textbf{Scenario} & \textbf{DM Scenario Description} \\
\hline
\endhead

\textbf{Scenario 1: Apartment Renting} & 
Alex is a 22-year-old incoming graduate student at XYZ University who is looking for a one-bedroom apartment off-campus. There are two types of housing options available, each with distinct characteristics and potential benefits. Alex has three main priorities for his decision: \textit{closer to campus}, \textit{affordability}, and \textit{safety}. With Alex’s priorities in mind, evaluate the two available rental options. After reviewing the information, think of yourself as Alex and, from his perspective, decide which rental choice would be the best fit. \\
\hline

\textbf{Scenario 2: Stock Investment} & 
Sam is a 30-year-old novice investor who is looking to allocate his savings into stocks. He came across two unique fictional stocks, each with its own set of characteristics. Sam has three main priorities for his investments: \textit{minimize risk}, \textit{steady growth}, and \textit{management quality}. With Sam’s priorities in mind, evaluate the two available stock options. After reviewing the information, think of yourself as Sam and, from his perspective, decide the best stock option for his investment. \\
\hline

\textbf{Scenario 3: Medication Treatment} & 
Emily is faced with a critical decision regarding the treatment of her 35-year-old cousin, Max, who has no known resistance to medications and no harmful habits. There are two types of medications available, each with distinct characteristics and potential benefits. Max has three main priorities for his treatment: \textit{minimize cost}, \textit{easy access}, and \textit{reputation and effectiveness}. With Max’s priorities in mind, evaluate the two available medication options. After reviewing the information, think of yourself as Emily and, from her perspective, decide and recommend the best treatment for Max. \\
\hline

\end{longtable}

\textbf{Norming Study for Importance Values.}  
After 2 researchers finished designing six information lists, we conducted a separate norming study to validate the importance values and reduce researcher bias. Following procedures used in prior work \citep{Rey2020}, we recruited 29 online participants through Prolific (university IRB approved). Participants completed a 45-minute survey after signing informed consent. All information pieces were presented in an anonymous Google Sheet, and participants rated each item on a 10-point scale (1 = very low influence, 10 = very strong influence) according to how much it would influence decision-making. To guide evaluation and reduce idiosyncratic biases, we provided a fixed persona with three prioritized factors (see Appendix Table~\ref{tab:study1_personas}). In addition, participants indicated their attitude toward each piece (positive, neutral, negative). Attention checks were included to ensure valid responses. After completing ratings for each scenario, participants selected a final choice (Option A or B) before proceeding to the next scenario.  

To compare population ratings with our initial researcher-defined importance values, we calculated Pearson correlations across the six information lists (2 options $\times$ 3 scenarios). Results showed moderate to strong correlations ($r = 0.41$–$0.65$; see Appendix Table~\ref{tab:correlation_results}), indicating good alignment between researcher-defined and participant-derived importance values. This provides strong validation for the use of self-defined values in the study.  

\textbf{Deriving the Best Option.}
\label{appendix:findbestoption}
Best options were determined using evaluation scores adapted from a prior work \citep{Rey2020}. For each information item, we first computed the mean importance rating across participants. For example, in the rental scenario, rental price received a high importance rating ($M=8.23$), while community noise received a lower rating ($M=4.13$). Next, we incorporated participants’ attitudes by applying a multiplier: $+1$ for positive, $-1$ for negative, and $0$ for neutral. Thus, a favorable rental price added $+8.23$ to the option score, while an unfavorable one subtracted $-8.23$. This process was repeated for all items, and adjusted scores were averaged to yield an evaluation score for each option.  

For example, across the three preliminary study scenarios (six options total), results indicated that Option A was preferred in Scenario 1 (Apartment Renting), while Option B was preferred in Scenarios 2 (Stock Investment) and 3 (Medication Treatment). These benchmark scores were later used to assess decision accuracy in the experiment.

\begin{table}[h]
\centering
\caption{Pearson Correlation on Importance Rating}
\label{tab:correlation_results}
\begin{tabular}{|c|c|c|c|}
\hline
\textbf{Option} & \textbf{Scenario 1} & \textbf{Scenario 2} & \textbf{Scenario 3} \\
 & \textit{Apartment Renting} & \textit{Stock Investment} & \textit{Medical Treatment Selection} \\
\hline
A & 0.5917 & 0.4315 & 0.4086 \\
B & 0.6495 & 0.5623 & 0.4159 \\
\hline
\end{tabular}
\end{table}

\subsubsection{Designing Information Pieces for Main Study}
\label{appendix:study2infodesign}
\begin{longtable}{|p{5cm}|p{9cm}|}
\caption{Main Study Decision-making (DM) Scenarios} \label{tab:study2_personas} \\
\hline
\textbf{Scenario} & \textbf{DM Scenario Description} \\
\hline
\endfirsthead

\hline
\textbf{Scenario} & \textbf{DM Scenario Description} \\
\hline
\endhead

\textbf{Scenario 1: Finding the Best Pet for Emily} & 
Emily is a 32-year-old software engineer who lives in a small modern home. She is seeking a pet that can keep her company. Two unique kinds of pets are available, each with distinct characteristics and potential benefits. With Emily's priorities in mind, evaluate the two available options. Consider how well each option meets her needs. After reviewing the information, think of yourself as Emily and, from her perspective, decide which choice would be the best fit. \\
\hline

\textbf{Scenario 2: Which Home Automation Device is More Preferred?} &
Jack is a 36-year-old surgeon who works extensive hours. He is seeking a tech-savvy home automation device that will serve as a practical solution to unburden his home life. With Jack's priorities in mind, evaluate the two available options. Consider how well each option meets his needs. After reviewing the information, think of yourself as Jack and, from his perspective, decide the best home device option for him. \\
\hline

\textbf{Scenario 3: Which is the Best Space Tourist Experience?} &
Marcus is a 28-year-old business owner. To celebrate an anniversary with his girlfriend, he is seeking a unique and thrilling tourist experience. Two attractions are available, each with distinct characteristics and potential benefits. With Marcus's priorities in mind, evaluate the two available options. Consider how well each option meets his needs. After reviewing the information, think of yourself as Marcus and, from his perspective, decide the best tourist option for him. \\
\hline

\end{longtable}

\section{Defining the Optimal Information Search Range}
\label{appendix:optimalrange}

In both the preliminary and main studies, we aimed to evaluate participants' decision-making efficiency by defining an \textit{optimal information search range} that distinguishes premature stopping, efficient search, and over-sampling behaviors. In the preliminary study, we pre-structured the presentation order of information pieces and their corresponding importance levels to examine whether participants would stop search at an optimal point. Specific stopping ranges were pre-defined to guide participants toward making informed decisions without over-sampling. Two trained junior researchers independently reviewed the ordered information lists to set these ranges, achieving high inter-rater reliability (ICC = .87); discrepancies were resolved through discussion. Participants were expected to stop search after reviewing between 8 and 16 pieces of information in the Decremental condition, and between 10 and 18 pieces in the Local Optimum (Local-Opt) condition. Random distribution had no predefined range due to sequence unpredictability. 

In the main study, the optimal range was refined using results from pilot testing and quantitative analyses of information importance values for each decision scenario. A fixed lower bound of four information pieces ensured minimal sufficiency across all conditions, while upper bounds were scenario- and distribution-specific, calibrated from the convergence points at which additional search no longer improved decision accuracy. Participants who sampled four or fewer information pieces were classified as \textit{Before} the optimal range, those search between five and the upper threshold as \textit{Within}, and those exceeding it as \textit{Slower}. Table~\ref{tab:optimalrange_short} summarizes the thresholds applied across scenarios and environmental uncertainty.

\begin{table}[ht]
\centering
\caption{Main Study: Scenario-specific optimal ranges.}
\label{tab:optimalrange_short}
\begin{tabular}{lcc}
\toprule
\textbf{Scenario} & \textbf{Decremental Range} & \textbf{Local-Opt Range} \\
\midrule
1 & 5--12 & 5--14 \\
2 & 5--14 & 5--16 \\
3 & 5--20 & 5--24 \\
\bottomrule
\end{tabular}
\end{table}

\section{Interface Design}

\subsection{Norming Study Display}
\begin{figure}[H]
    \centering
    \includegraphics[width=\linewidth]{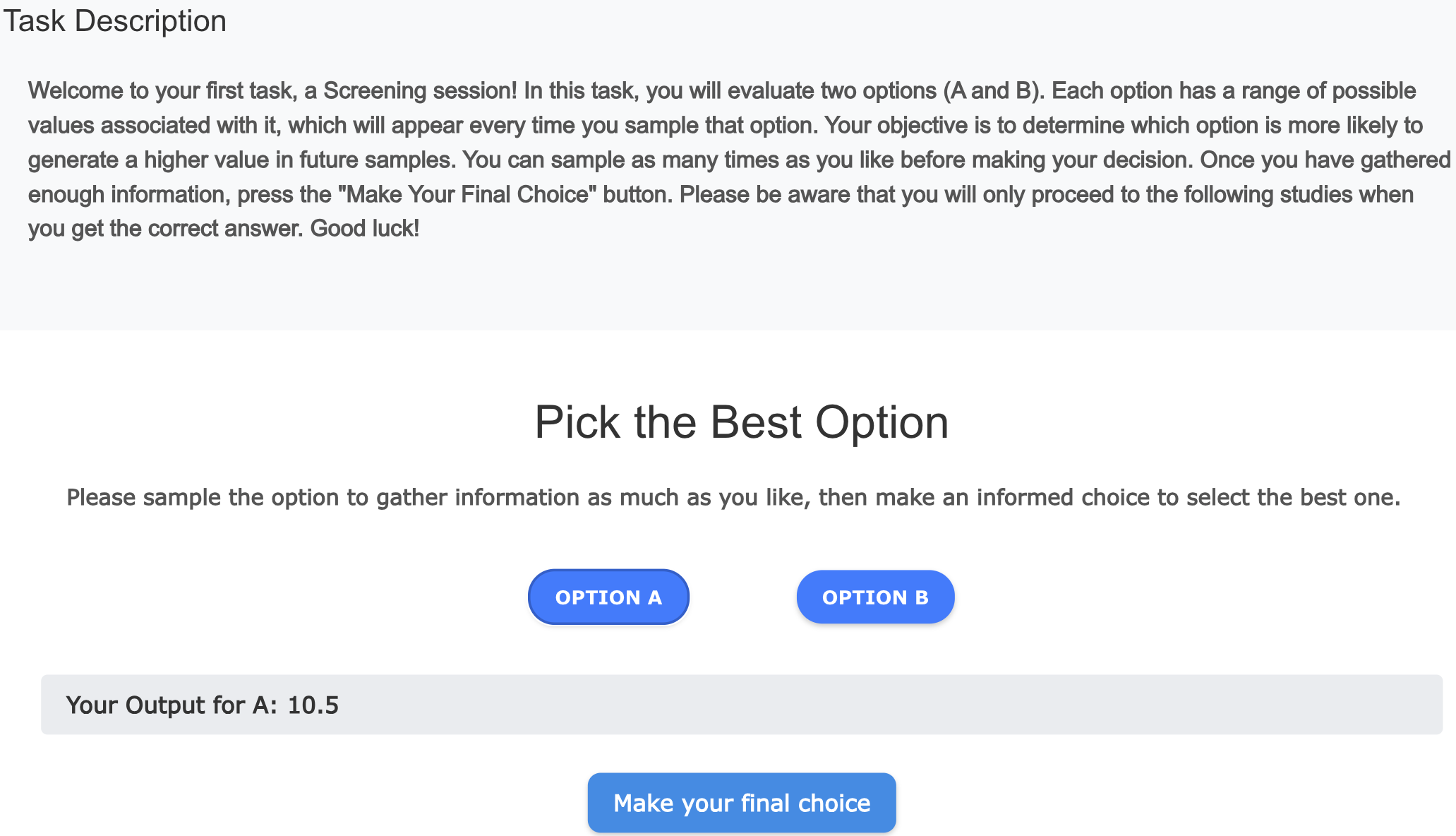}
    \caption{Screenshot of the Norming Study (search stage)}
    \label{fig:screening}
\end{figure}

\begin{figure}[H]
    \centering
    \includegraphics[width=\linewidth]{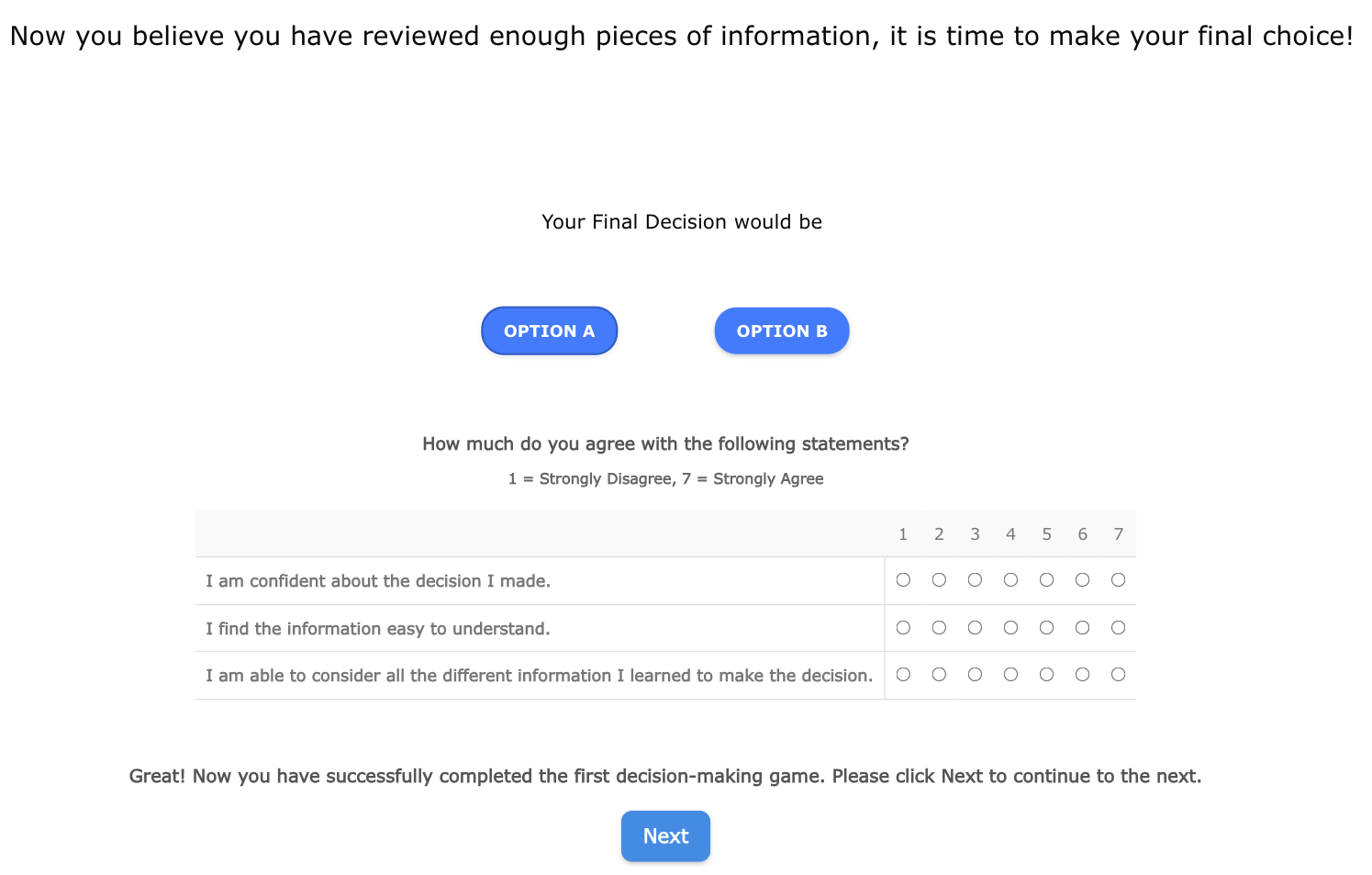}
    \caption{Screenshot of the Norming Study (choice stage)}
    \label{fig:screening_choice}
\end{figure}

\begin{figure}[H]
    \centering
    \includegraphics[width=\linewidth]{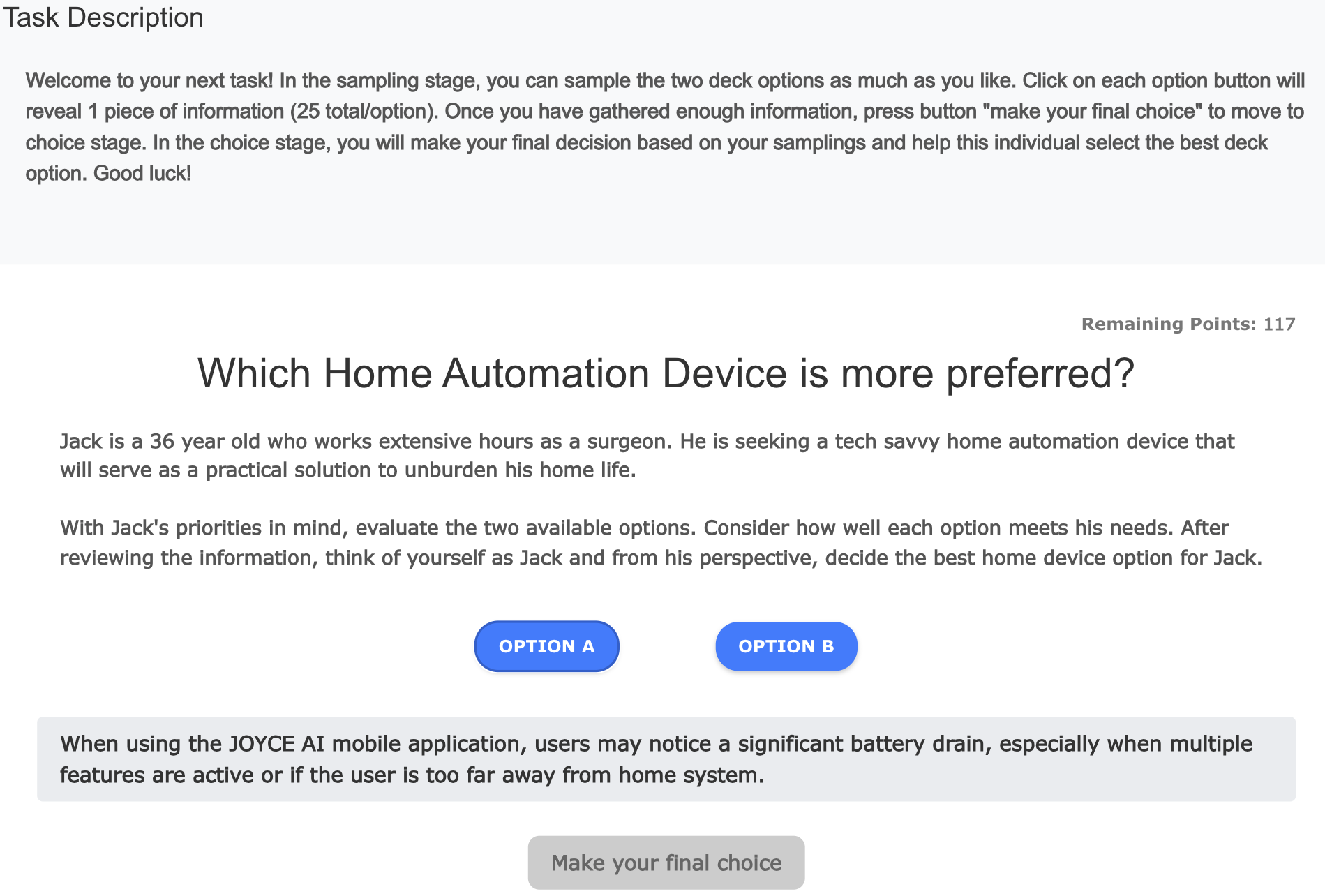}
    \caption{Screenshot of the Main Study DM Scenario in No-SERA condition (search stage)}
    \label{fig:study2_nosera}
\end{figure}

\end{document}